\documentclass[twocolumn,amsfont,amssymb,amsmath, showpacs,balancelastpage, nofootinbib]{revtex4-1}
\pdfoutput=1

\usepackage{graphicx}
\usepackage{dcolumn}
\usepackage{bm}
\usepackage{amssymb,amsmath,bm}  
\usepackage{color}
\usepackage[colorlinks,linkcolor=red,citecolor=blue,urlcolor=blue ]{hyperref}
\usepackage{multirow}
\usepackage[utf8]{inputenc}
\usepackage{balance}
\usepackage{enumitem}
\usepackage{lipsum}
\newcommand{\nv}{\hat{\bf n}}
\newcommand{\kalo}{Karhunen-Lo\`{e}ve\,}
\newcommand{\jcap}{JCAP}
\newcommand{\mnras}{MNRAS}
\newcommand{\aap}{A\&A}

\newcommand{\apjs}{ApJS}
\newcommand{\apjl}{ApJL}

\newcommand{\procspie}{Proceedings of the SPIE}
\newcommand{\physrep}{Physics Reports}
\begin{document}
\title{Science-driven 3D data compression}
\author{David Alonso$^1$}
\affiliation{$^{1}$Oxford Astrophysics, Department of Physics, Keble Road, Oxford, OX1 3RH, UK}

\begin{abstract}
  Photometric redshift surveys map the distribution of matter in the Universe through the positions and shapes of galaxies with poorly resolved measurements of their radial coordinates. While a tomographic analysis can be used to recover some of the large-scale radial modes present in the data, this approach suffers from a number of practical shortcomings, and the criteria to decide on a particular binning scheme are commonly blind to the ultimate science goals. We present a method designed to separate and compress the data into a small number of uncorrelated radial modes, circumventing some of the problems of standard tomographic analyses. The method is based on the \kalo transform (KL), and is connected to other 3D data compression bases advocated in the literature, such as the Fourier-Bessel decomposition. We apply this method to both weak lensing and galaxy clustering. In the case of galaxy clustering, we show that the resulting optimal basis is closely associated with the Fourier-Bessel basis, and that for certain observables, such as the effects of magnification bias or primordial non-Gaussianity, the bulk of the signal can be compressed into a small number of modes. In the case of weak lensing we show that the method is able to compress the vast majority of the signal-to-noise\`{A} into a single mode, and that optimal cosmological constraints can be obtained considering only three uncorrelated KL eigenmodes, considerably simplifying the analysis with respect to a traditional tomographic approach.
\end{abstract}

\maketitle

\section{Introduction}\label{sec:intro}
  Astronomical data is inherently three-dimensional: the main observable is the intensity of the electromagnetic emission in the sky as a function of wavelength and line-of-sight direction, determined by two angles. An idealized cosmological analysis would therefore use, as a data vector, the full cube $I(\lambda,\theta,\phi)$ probed over sufficiently well resolved angular and frequency scales \cite{2014arXiv1403.3727D}. However, the operational costs of obtaining such a dataset imply that we can realistically only access a compressed version of it, where the compression method comes in different flavours:
  \begin{itemize}
    \item We can decrease the measurement noise by integrating the sky intensity over large frequency bands. This approach has been used for instance in CMB observations \cite{2013ApJS..208...20B,2016A&A...594A...1P,2016arXiv161002743A}.
    \item Angular resolution can also be sacrificed for wider sky and frequency coverage, as has been proposed for future intensity mapping experiments \cite{2004MNRAS.355.1339B,2014SPIE.9145E..22B,2015aska.confE..19S}.
    \item The size of the dataset can also be reduced by collecting only the flux associated with the brightest extra-galactic objects. An accurate measurement of their spectra then allows a determination of their redshift, producing the well-known spectroscopic redshift surveys \cite{2017MNRAS.470.2617A,2013arXiv1308.0847L,2011arXiv1110.3193L}.
    \item Measuring individual spectra is a costly operation, however, and can usually only be done for a small subsample of all available sources. This problem can be mitigated by inferring the source's redshift from their emission in a small number of wider frequency bands, in what is known as a photometric redshift survey \cite{2016arXiv161205560C,2016MNRAS.460.1270D,2017arXiv170208449A}. The redshifts thus determined are far less precise than their spectroscopic counterparts, and usually only an imperfect estimation of the redshift probability distribution for each galaxy is accessible.
  \end{itemize}

  Even after this first compression stage, the size of the dataset makes a direct analysis of it as a data vector a computationally intractable problem. Typically this should not be an issue in terms of information loss, since large portions of the data are usually dominated by measurement noise, contaminated by sources of systematic uncertainty (both observational and theoretical) or contain only redundant information. An efficient data compression method will therefore identify these sections of data space and eliminate them, or collect them into summary statistics, while minimizing the loss of meaningful cosmological information. An example of this is the standard analysis of cosmological datasets in terms of their two-point statistics \cite{1997PhRvD..55.5895T}. However, even in this case the resulting data vector can be large enough to present an important computational challenge in terms of likelihood evaluation and covariance estimation. The complexity of latter problem, in particular, scales with the square of the data vector size, and can become an important drain of computational resources \cite{2013PhRvD..88f3537D,2017arXiv170706529H}.
  
  In this work we will concern ourselves with the topic of 3D data compression: the problem of identifying the uncorrelated angular and radial modes of the data that optimally contain the maximum amount of information. This problem has been previously addressed in the literature \cite{2003MNRAS.343.1327H,2013arXiv1307.1307M,2015PhRvD..92l3010L,2014arXiv1403.5553K}, and a number of approaches have been proposed depending on the definition of uncorrelatedness used, and on the type of information one wishes to preserve. Here we will present a method to derive a set of uncorrelated radial eigenmodes that are manifestly optimal in terms of information compression for any quantity, such as individual cosmological parameters or the amplitude of the cosmological signal over any set of known contaminants. The method is based on the well-understood \kalo transform \cite{1997ApJ...480...22T,1996ApJ...465...34V}, and is similar in spirit to the derivation of optimal weighting schemes for the analysis of spectroscopic surveys \cite{1994ApJ...426...23F,2017arXiv170205088M}. Although we will focus here on the case of photometric redshift surveys, the method can be applied to any set of cosmological datasets.
  
  The article is structured as follows: Section \ref{sec:method} describes the \kalo (KL hereupon) transform and its applicability in the context of 3D data compression. Section \ref{sec:results} shows the performance of the method in a number of science cases, such as the derivation of optimal radial bases for galaxy clustering (Sec. \ref{ssec:results.gc}) and weak lensing (Sec. \ref{ssec:results.wl}) observations and the use of the KL eigenmodes to measure the effects of primordial non-Gaussianity (Sec. \ref{ssec:results.fnl}) and lensing magnification (Sec. \ref{ssec:results.mag}) with a small number of modes. Finally Section \ref{sec:discussion} summarizes our findings and dicusses the advantages and shortcomings of the method.

\section{Method}\label{sec:method}
  \subsection{The Karhunen-Loeve transform}\label{ssec:method.klbasis}
    The idea behind the \kalo transform, as developed within the field of cosmological data analysis in e.g. \cite{1996ApJ...465...34V,1997ApJ...480...22T}, is to compress a given data vector into a small set of modes containing most of the useful information on a particular parameter (or set of parameters). Let ${\bf x}$ be a data vector of dimension $N_s$, and let $\theta$ be a particular parameter we want to measure. Under the assumption that ${\bf x}$ is Gaussianly distributed with mean 0 and covariance ${\sf C}$, a set of linear combinations $y_p\equiv{\bf e}_p^\dag\,{\bf x}$ can be found such that the $y_p$ are white and uncorrelated ($\langle y_py_q^*\rangle=\delta_{pq}$), and such that the first $m<N_s$ combinations contain most of the information about $\theta$. This is done by solving the generalized eigenvalue problem \cite{1997ApJ...480...22T}:
    \begin{equation}\label{eq:kl_general}
      \partial_\theta{\sf C}\,{\bf e}_p=\lambda_p\,{\sf C}\,{\bf e}_p,
    \end{equation}
    where $\partial_\theta\equiv\partial/\partial_\theta$.
    
    Although the \kalo transform can be used to compress the information on any particular parameter, it has been most commonly used to separate signal-dominated and noise-dominated modes by optimizing for the amplitude of the signal, as we explore below. Before moving on, however, it is worth noting that a generalized eigenvalue problem such as Eq. \ref{eq:kl_general} can always be recast as a standard eigenvalue problem of the form ${\sf A}\,\tilde{\bf e}_p=\lambda_p\,\tilde{\bf e}_p$, where
    \begin{equation}
      {\sf A}\equiv {\sf C}^{-1/2}\,(\partial_\theta{\sf C})\,{\sf C}^{-1/2},\hspace{12pt}
      \tilde{\bf e}_p\equiv{\sf C}^{1/2}{\bf e}_p,
    \end{equation}
    and we have made use of the fact that ${\sf C}$ is positive-definite (and therefore ${\sf C}^{1/2}$ is well-defined and invertible).

    \subsubsection{The KL transform for the signal-to-noise}\label{sssec:method.klbasis.sn}
      Let us decompose the data vector ${\bf x}$ into uncorrelated signal and noise components ${\bf x}={\bf s}+{\bf n}$ where, in this context, the signal is the part of the data containing any information of cosmological interest, and the noise is any contaminant preventing us from accessing it\footnote{${\bf n}$ could include, for instance, the contribution of foregrounds in intensity mapping experiments, which motivates the use of the KL transform as a foreground cleaning method \cite{2014ApJ...781...57S}.}. In this particular case, the data covariance matrix can be split into their independent contributions ${\sf C}={\sf S}+{\sf N}$.
      
      The KL transform has traditionally been used to design an eigenbasis that maximizes the overall signal-to-noise ratio (e.g \cite{1995PhRvL..74.4369B,1996ApJ...465...34V}). This can be done by defining a fictitious parameter $\rho$ multiplying the signal part of the data with fiducial value $\rho=1$ (i.e. ${\bf x}=\rho{\bf s}+{\bf n}$). In this case, after some trivial manipulations, the eigenvalue equation (Eq. \ref{eq:kl_general}) takes the form:
      \begin{equation}\label{eq:kl_sn}
        ({\sf S}+{\sf N}){\bf e}_p=\lambda_p{\sf N}{\bf e}_p,
      \end{equation}
      where we have redefined $2/(2-\lambda_p)\rightarrow\lambda_p$. This can be cast into a standard eigenvalue equation using the Cholesky decomposition of the noise covariance matrix ${\sf N}={\sf L}{\sf L}^\dag$:
      \begin{equation}\label{eq:kl_sn_st}
        \left[{\sf L}^{-1}{\sf C}\,({\sf L}^{-1})^\dag\right]\,\tilde{\bf e}_p=\lambda_p\tilde{\bf e}_p,
      \end{equation}
      where $\tilde{\bf e}_p\equiv{\sf L}^\dag{\bf e}_p$.
      
      At this point it is worth noting that the generalized eigenvalue problem in Eq. \ref{eq:kl_sn} can be understood as the problem of diagonalizing ${\sf C}$ under a non-standard dot product $\circ$ given by the inverse noise covariance matrix (i.e. ${\bf a}\circ{\bf b}\equiv{\bf a}^\dag{\sf N}^{-1}{\bf b}$). Under this dot product, an eigenbasis ${\sf F}\equiv({\bf f}_1,{\bf f}_2,...,{\bf f}_{N_s})$ can be found such that ${\sf F}$ is orthonormal ${\sf F}\circ{\sf F}={\sf I}$, and the covariance of the transformed data vector ${\bf y}\equiv{\sf F}\circ{\bf x}$ is diagonal:
      \begin{equation}\label{eq:kl_sn_prod}
        \langle{\bf y}\,{\bf y}^\dag\rangle={\sf F}^\dag{\sf N}^{-1}{\sf C}{\sf N}^{-1}{\sf F}={\sf \Lambda}\equiv{\rm diag}(\lambda_1,...,\lambda_{N_s}).
      \end{equation}
      Using the orthonormality of ${\sf F}$ (with respect to the non-standard dot product), this can be cast into the same form as Eq. \ref{eq:kl_sn_st}, where ${\bf f}_p={\sf L}\tilde{\bf e}_p={\sf N}{\bf e}_p$.
      
      Finally, note that, because both ${\sf S}$ and ${\sf N}$ are positive-definite matrices, their eigenvalues will also be positive. Since the eigenvalues of ${\sf N}$ under the KL transform are, by construction, 1, the elements of ${\sf \Lambda}$ above will all be greater than 1, and converging to 1 for the noise-dominated modes.
      
    \subsubsection{The KL transform with correlated contaminants}\label{sssec:method.klbasis.cr}
      Let us now consider a more general case in which we further split the noise into two parts ${\bf n}\rightarrow{\bf n}+{\bf m}$, where ${\bf m}$ is a contaminant with a non-zero correlation with the signal. The covariance matrix of the data is then given by:
      \begin{equation}
        \langle{\bf x}\,{\bf x}^\dag\rangle=\rho^2{\sf S}+2\rho{\sf M}_s+{\sf M}+{\sf N},
      \end{equation}
      where ${\sf M}_s\equiv(\langle{\bf m}\,{\bf s}^\dag\rangle+\langle{\bf s}\,{\bf m}^\dag\rangle)/2$, ${\sf M}\equiv\langle{\bf m}\,{\bf m}^\dag\rangle$ and we have kept the fictitious parameter $\rho$ defined in the previous section.
      Eq. \ref{eq:kl_general} then reads:
      \begin{equation}\label{eq:kl_sn_corr}
        \left({\sf S}+{\sf M}_s\right)\,{\bf e}_p=
        \frac{\lambda_p}{2}{\sf C}\,{\bf e}_p.
      \end{equation}
      
      Unfortunately, in this case the manipulation that lead us to Eq. \ref{eq:kl_sn} cannot be performed. If we were to do so, the matrix remaining on the right hand side of this equation would not be positive-definite, and the corresponding generalized eigenvalue problem would be ill-defined. This is not a problem, since the solutions to Eq. \ref{eq:kl_sn_corr} still separate the modes with the highest signal. The separation of the noise-dominated modes becomes less obvious, however, since the resulting eigenvalues cannot be simply compared with 1, corresponding to noise-dominated modes in the previous section.
      
      The eigenvector solutions to the generalized eigenvalue problem in Eq. \ref{eq:kl_sn_corr} can be collected as columns of a matrix ${\sf E}$ that simultaneously satisfies the equations:
      \begin{equation}
        {\sf E}^\dag\left({\sf S}+{\sf M}_s\right){\sf E}={\sf \Lambda},
        \hspace{12pt}
        {\sf E}^\dag{\sf C}{\sf E}={\sf I},
      \end{equation}
      where ${\sf I}$ is the identity and ${\sf \Lambda}={\rm diag}(\lambda_1,...,\lambda_{N_s})$.
      Since the second equation implies ${\sf C}{\sf E}\equiv({\sf E}^\dag)^{-1}$, the original vector ${\bf x}$ can be recovered from the coefficients ${\bf y}\equiv(y_1,...,y_{N_s})$ as ${\bf x}={\sf C}\,{\sf E}\,{\bf y}$. More interestingly, one can identify the principal eigenvectors of the Eq. \ref{eq:kl_sn_corr} (e.g. those with associated eigenvalues $\lambda_p$ above a given threshold $\lambda_{\rm thr}$) and project out the remaining modes, which are presumably more contaminated by ${\bf m}$. This procedure defines a filter ${\sf W}\equiv{\sf C}\,{\sf E}\,{\sf P}\,{\sf E}^\dag$, where ${\sf P}$ is a projection matrix with $1$s in the diagonal elements corresponding to the principal eigenmodes and zeros everywhere else. The filtered data vector is therefore $\tilde{\bf x}={\sf W}{\bf x}$.

  \subsection{Application to tomographic datasets}\label{ssec:method.tomographic}
    The standard method to draw cosmological constraints from photometric redshift surveys is to divide the galaxy sample into bins in photo-$z$ space and use the information encoded in all the relevant auto- and cross-correlations between different bins \cite{2012MNRAS.427.1891A,2016PhRvD..94b2002B,2017MNRAS.465.1454H}, making use of various calibration methods in order to estimate the true redshift distribution of each bin. Several criteria can be followed in order to select these redshift bins, such as minimizing the correlation between non-neighbouring bins or preserving a roughly constant number density on all bins. Other approaches
    (\cite{2003MNRAS.343.1327H,2007MNRAS.376..771K,2014MNRAS.442.1326K})
    involve projecting the main observable (e.g. galaxy overdensity or shear) onto the Fourier-Bessel eigenbasis. None of these schemes are manifestly optimal from the point of view of $S/N$, final cosmological constraints or contaminant deprojection, however. This section presents an alternative slicing scheme addressing these shortcomings, based on the KL transform.
    
    \subsubsection{Tomographic analyses}\label{sssec:method.tomographic.st}
      Let us start by assuming that we have split the galaxy sample into $N_s$ subsamples. As mentioned above, we will think of each of these subamples as some kind of redshift binning (e.g. binning galaxies in terms of their maximum-likelihood redshift), but the formalism applies to any set of subsamples. Let $a^\alpha(\nv)$ be the a field on the sphere at the angular position $\nv$ and defined in terms of the properties of the sources in the $\alpha$-th sample (e.g. the cosmic shear field $\gamma^\alpha$ or the galaxy overdensity $\delta^\alpha$), and let $\phi^\alpha(z)$ be the redshift distribution of these sources. Finally, let $a^\alpha_{\ell m}$ be the spherical harmonic coefficients of $a^\alpha$\footnote{Spin-2 fields, such as the cosmic shear, will be decomposed in spin-2 spherical harmonics, however the discussion below holds for fields of arbitrary spin.}. The power spectrum for our set of subsamples is defined as the two-point correlator of $a^\alpha_{\ell m}$:
      \begin{equation}
        \left\langle {\bf a}_{\ell m}\,{\bf a}^\dag_{\ell' m'} \right\rangle\equiv\delta_{\ell\ell'}\delta_{mm'}{\sf C}_\ell,
      \end{equation}
      where we have packaged $a^\alpha_{\ell m}$ as a vector for each $(\ell,m)$: ${\bf a}_{\ell m}\equiv(a^1_{\ell m},...,a^{N_s}_{\ell m})$. Usually the observed field can be decomposed into uncorrelated signal and noise components ${\bf a}={\bf s}+{\bf n}$, with a similar decomposition in the power spectrum, ${\sf C}_\ell={\sf S}_\ell+{\sf N}_\ell$.
  
      Once the choice of subsamples $\alpha$ is made, the standard analysis method would proceed by performing a likelihood evaluation of the two-point statistics of these subsamples. While this procedure is relatively simple, it suffers from a number of drawbacks, an incomplete list of which is:
      \begin{enumerate}
        \item It is not clear what the optimal strategy should be to define the sub-samples. The brute-force solution to make sure one exploits all of the information present in the data would be to use a large number of very narrow redshift bins, and let the likelihood evaluation pick up the information encoded in them.
        \item $C^{\alpha\beta}_\ell$ is a $N_s\times N_s\times N_\ell$ data vector. Thus increasing $N_s$ will increase the computational time required for each likelihood evaluation like $N_s^2$ and number of elements of the covariance matrix of $C^{\alpha\beta}_\ell$ like $N_s^4$, with the corresponding increase in complexity needed to estimate this covariance. Although this can be partially alleviated by considering only correlations between neighbouring redshift shells, the amount of information lost by neglecting all correlations beyond a given neighbouring order is not clear a priori.
        \item Estimating the redshift distribution for a large number of subsamples can be inaccurate, depending on the method used to do so, on the quality of the photometric redshift posterior information and on the statistics of the available spectroscopic sample.
     \end{enumerate}
  
    \subsubsection{Optimal radial eigenbasis}\label{sssec:method.tomographic.kl}
      Following the description in Section \ref{sssec:method.klbasis.sn}, it is straightforward to derive an optimal set of radial, uncorrelated eigenmodes.
      \begin{enumerate}
        \item We start by assuming that the field ${\bf a}$ has been measured in a number of narrow redshift bins, and by defining the inverse-variance weighted field $\tilde{\bf a}_{\ell m}\equiv{\sf N}^{-1}_\ell\,{\bf a}_{\ell m}$.
        \item Let us consider a set of linear combinations of the weighted field measured on narrow redshift bins:
        \begin{equation}
          {\bf b}_{\ell m}={\sf F}_\ell^\dag\cdot\tilde{\bf a}_{\ell m}\equiv{\sf F}_\ell\circ{\bf a},
        \end{equation}
        where ${\sf F}_\ell$ is a yet-unspecified matrix and, as in Section \ref{sssec:method.klbasis.sn}, we have let ${\sf N}^{-1}_\ell$ define the non-standard dot product ${\bf v}_\ell\circ{\bf w}_\ell\equiv{\bf v}^\dag_\ell\cdot{\sf N}^{-1}_\ell\cdot{\bf w}_\ell$. The power spectrum for this new observable would then simply be given by:
        \begin{equation}\label{eq:dp_uncouple}
          {\sf D}_\ell\equiv\left\langle{\bf b}_{\ell m}\,{\bf b}^\dag_{\ell m}\right\rangle={\sf F}_\ell^\dag\circ{\sf C}_\ell\circ{\sf F}_\ell.
        \end{equation}
        \item Requiring that the new modes be uncorrelated, we can identify Eq. \ref{eq:dp_uncouple} with the generalized eigenvalue equation \ref{eq:kl_sn_prod}, which defines the KL eigenbasis ${\sf F}_\ell$ by additionally requiring that it be orthonormal (${\sf F}_\ell\circ{\sf F}_\ell={\sf I}$). Note that, after this transformation and without any further optimization, some of the practicalities of the original problem are already simplified, since we can now focus on the diagonal elements of the new power spectrum and its covariance.
        \item The data can be further compressed by assuming that we are interested in measuring a set of cosmological parameters $\Theta\equiv\{\theta_1,...\}$. The information regarding this set of parameters encoded in a given data vector ${\bf x}$ can be quantified in terms of its Fisher matrix (the expectation value of the Hessian of the log-likelihood with respect to $\Theta$), which assuming $\langle{\bf x}\rangle=0$ reads
        \begin{equation}
          {\cal F}_{ij}\equiv\left\langle\partial_i\partial_j{\cal L}\right\rangle=\frac{1}{2}{\rm Tr}\left(\partial_i{\sf C}\,{\sf C}^{-1}\partial_j{\sf C}\,{\sf C}^{-1}\right),
        \end{equation}
        where ${\sf C}\equiv\langle {\bf x}\,{\bf x}^\dag\rangle$ is the covariance matrix of the data.
        We can thus rank the eigenvectors $({\sf F}_\ell)^p_\alpha$ in terms of their information content (in a Fisher-matrix sense). In the simplest scenario one may be interested in maximizing the overall signal-to-noise ($S/N$), in which case each mode contributes independently to the Fisher matrix element of the signal amplitude.
        \item The final set of uncorrelated modes can then be truncated to the first $M$ defined by this procedure, which will contain the bulk of the information needed to constrain $\Theta$.
    \end{enumerate}
    Besides the elegance of this method in defining a natural set of radial basis functions for the particular dataset under study, analogous to the Fourier-Bessel basis in a translationally-invariant system (see Section \ref{ssec:results.bessel}), its merits are better evaluated in terms of data compression. This strategy allows one to reliably and significantly reduce the dimensionality of the data vector from $N_s^2\times N_\ell$ to $M\times N_\ell$ while minimizing the loss of information. This can lead, for instance, to a substantial reduction of the computational costs of likelihood sampling and covariance estimation.
    
    Note that, although the method is based on an initial thin-slicing of the galaxy distribution, the fact that the final datased comprises only a small set of samples means that the method is not penalized in terms of photometric redshift uncertainties. Once the KL eigenmodes ${\sf F}_\ell$ are found for a fiducial cosmological model, they can be directly applied as weights to all the objects in the survey to generate the $b^p$ modes. These modes are be characterized by their own window function:
    \begin{equation}\label{eq:window_kl}
     \tilde{\phi}_\ell^p(z)=\sum_\alpha \frac{({\sf F}_\ell)^p_\alpha\,\phi^\alpha(z)}{N^{\alpha\alpha}_\ell},
    \end{equation}
    where we have assumed a diagonal noise power spectrum for simplicity.
    The same methods used to calibrate photo-$z$ uncertainties in the standard tomographic analysis can be applied on $b^p$ to calibrate $\tilde{\phi}^p$ with minor modifications (e.g. weighed and $\ell$-dependent stacking of photo-$z$ pdfs, or cross-correlations of the $b^p$ maps with a spectroscopic survey in the case of clustering redshifts). Furthermore, using ${\sf F}_\ell$ for the fiducial cosmology as model-agnostic weights and inserting them in Eq. \ref{eq:dp_uncouple}, the theoretical prediction for the power spectrum of each mode $D^p_\ell$ can be computed in a model-independent way.
    
    Finally, the method outlined in this section is based on the KL decomposition that maximizes the amplitude of the signal under study. This is the main application advocated in this article, since it is plausible that the set of modes containing the bulk of the cosmological signal will also drive the constraints on any comprehensive set of cosmological parameters (we explore this in more detail in Section \ref{ssec:results.wl}). However, we must note that, for individual parameters, the optimal degree of data compression is achieved by solving the general KL eigenvalue problem (\ref{eq:kl_general}), which can lead to substantial improvements with respect to the $S/N$-optimal basis. We explore one particular example of this in Section \ref{ssec:results.fnl}.

\section{Performance and particular examples}\label{sec:results}
  This Section explores the performance of the KL decomposition in a number of specific science cases.
  
  \subsection{Special case: the harmonic-Bessel basis}\label{ssec:results.bessel}
    Let us start by considering a simplified case where the field $a$ is the overdensity field of a non-evolving galaxy population for which we neglect the effects of redshift-space distortions. Let us further assume that we have perfect redshift information, such that we can split the sample into thin radial slices of equal width $\delta \chi$, which we label by their comoving radius $\chi$. The noise in the measurement of $a$ is given purely by shot noise, and since (as per our initial assumptions) the number density of sources does not change with $\chi$, the noise power spectrum is diagonal and scales like $N_\ell(\chi,\chi')\propto \delta_{\chi,\chi'}\,\chi^{-2}$. Thus, the dot product is just given by:
    \begin{equation}
      {\bf b}^\dag\circ{\bf c}\propto\int d\chi\,\chi^2\,b(\chi)^*\,c(\chi).
    \end{equation}
   
    In this case, the cross-shell signal power spectrum is given by,
    \begin{equation}
      S_\ell^{\chi\chi'}=\frac{2}{\pi}\int_0^\infty dk\,k^2\,P_k\,j_\ell(k\chi)j_\ell(k\chi'),
    \end{equation}
    and it is trivial to show that the KL eigenmodes are simply given by the spherical Bessel functions: $({\sf F}_\ell)^k_\chi\propto \sqrt{2/\pi}j_\ell(k\chi)$:
    \begin{widetext}
    \begin{align}\nonumber
      D_\ell^{kk'}&\equiv\sum_{\chi,\chi'}\left(F_\ell\right)^{k}_\chi\left(F_\ell\right)^{k'}_{\chi'}S_\ell^{\chi\chi'}\\\nonumber
      &\propto\frac{2}{\pi}\int d\chi\,\chi^2\int d\chi'\,\chi'^2 j_\ell(k\chi)j_\ell(k'\chi') S_\ell^{\chi\chi'}\\\nonumber
      &=\int dq\,q^2P_q\left[\frac{2}{\pi}\int d\chi\,\chi^2\,j_\ell(q\chi)j_\ell(k\chi)\right]\left[\frac{2}{\pi}\int d\chi'\,\chi'^2\,j_\ell(q\chi')j_\ell(k'\chi')\right]\\
      &=\int dq\,q^2P_q\frac{\delta(k-q)}{q^2}\frac{\delta(k'-q)}{q^2}=P_k\frac{\delta(k-k')}{k^2}=\frac{P_k}{k^2\Delta k}\delta_{k,k'}
    \end{align}
    \end{widetext}
   
    This choice of basis defines the so-called harmonic-Bessel (or Fourier-Bessel) decomposition, and has been used as a data-compression method for the analysis of photometric redshift datasets (e.g. \cite{2014MNRAS.442.1326K}). In any realistic scenario -- e.g. in the presence of redshift uncertainties, redshift-space distortions or in the analysis of weak lensing data -- this basis is non-optimal (among other things different $k$-modes will be correlated), as opposed to the KL basis described in the previous section.
    
  \subsection{Galaxy clustering - Bessel-like eigenfunctions}\label{ssec:results.gc}
    \begin{figure*}
      \centering
      \includegraphics[width=0.49\textwidth]{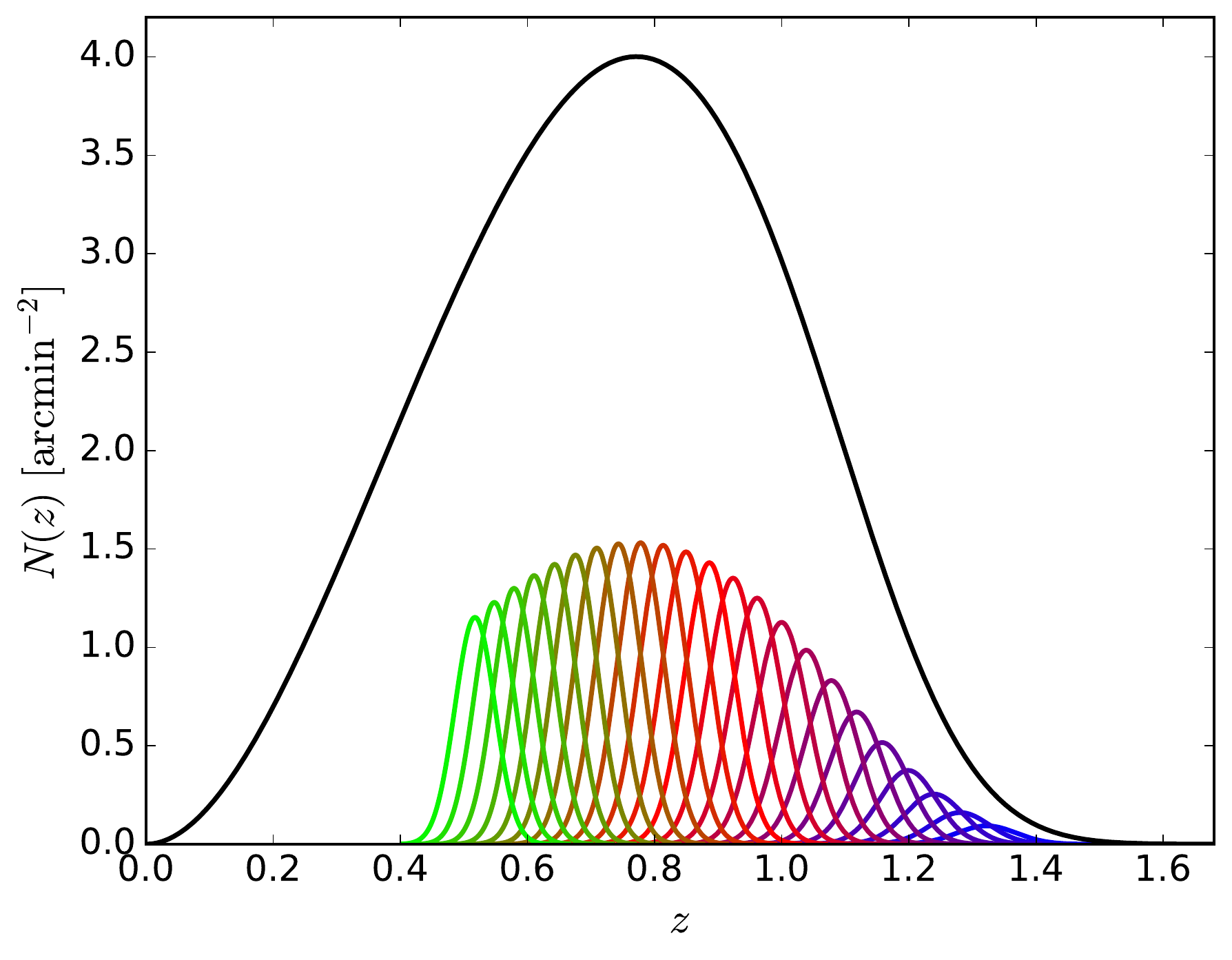}
      \includegraphics[width=0.49\textwidth]{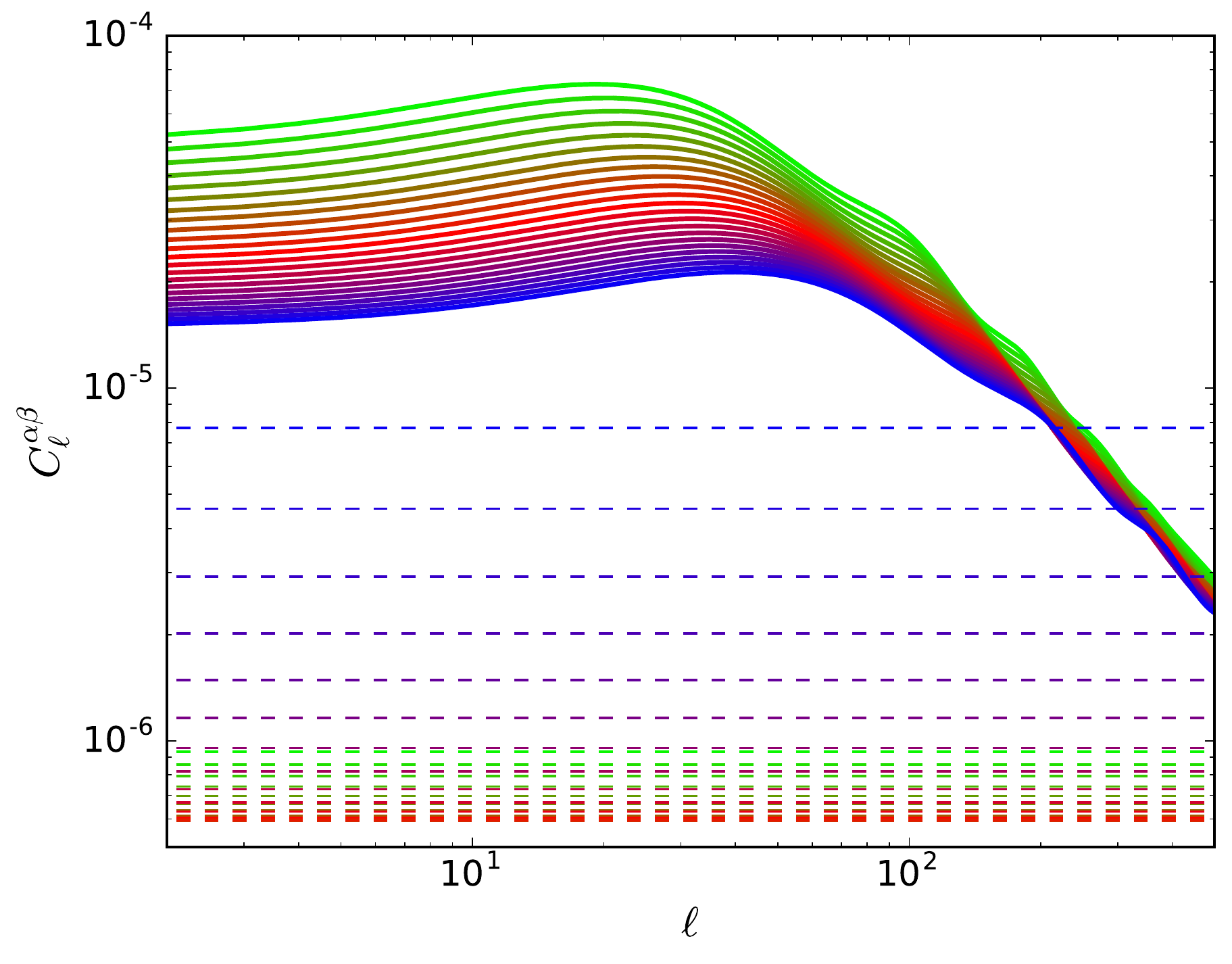}
      \caption{{\sl Left}: redshift distribution and bins considered for the KL analysis of a strawman large-scale-structure survey targeting a sample of red galaxies. {\sl Right}: clustering auto-power spectra of the redshift bins shown in the left panel. The signal and noise power spectra are shown as thick solid and thin dashed lines respectively.}\label{fig:nz_gc}
    \end{figure*}
    The assumptions used in the previous section are an ideal limit of the data collected by a photometric survey. In a more realistic (although still idealized) scenario, the information about the radial position of a given source is encoded in its posterior photo-$z$ distribution $p(z|\alpha)$, where $\alpha$ is a continuous variable determining the properties of the photo-$z$ (e.g. the mean of the posterior). The cross-power spectrum of two samples with photo-$z$ properties $\alpha$ and $\beta$ is given by
    \begin{align}
      &C_\ell^{\alpha\beta}=S^{\alpha\beta}_\ell+N^{\alpha\beta}_\ell,\\\label{eq:cl_generic}
      &S_\ell^{\alpha\beta}=\frac{2}{\pi}\int_0^\infty dk\,k^2\,\Delta_\ell^\alpha(k)\,\Delta_\ell^\beta(k),\\
      &N_\ell^{\alpha\beta}=\frac{\delta(\alpha-\beta)}{n_t\,p(\alpha)},
    \end{align}
    where $n_t$ is the total angular number density of sources and
    \begin{align}\nonumber
      &\Delta^{\alpha}_\ell(k)\equiv\int dz\,p(z|\alpha)\,\Psi_\ell(k,z)\,\sqrt{P(k,z)},\\\label{eq:tgc_dr}
      &\Psi_\ell(k,z)=b^\alpha(z)j_\ell(k\,\chi(z))-f(z)j_\ell''(k\,\chi(z)).
    \end{align}
    Here $b^\alpha(z)$ is the linear galaxy bias, $f(z)=d\log\delta/d\log a$ is the growth rate of structure, $P(k,z)$ is the matter power spectrum at redshift $z$, $p(\alpha)$ is the probability that a source has photo-$z$ properties $\alpha$, and $p(z|\alpha)$ is the conditional redshift distribution of these sources (we have labelled this quantity $\phi^\alpha(z)$ in previous sections). Note that, for simplicity, we have kept the contribution of redshift-space distortions at linear order and neglected the effect of magnification (this will be studied in Section \ref{ssec:results.mag}).
    
    For a continuous variable $\alpha$, the generalized eigenvalue problem in Eq. \ref{eq:kl_sn} becomes a homogeneous Fredholm integral equation of the second kind:
    \begin{align}
      \int d\beta\,C^{\alpha\beta}_\ell e^p_\ell(\beta)=\lambda_p\int d\beta\,N^{\alpha\beta}_\ell e^p_\ell(\beta)\Rightarrow\\
      \Rightarrow\int d\beta\,n_t\,p(\alpha) S^{\alpha\beta}_\ell\,e^p_\ell(\beta)=(\lambda_p-1)e^p_\ell(\alpha).
    \end{align}
    In the limit of perfect photo-$z$s ($p(z|\alpha)=\delta(z-\alpha)$), and in the absence of redshift-space distortions, the solution to this equation are the spherical Bessel functions, as proven in the previous section. For general kernels, however, no analytical solution to the homogeneous Fredholm equation can usually be found, and the standard procedure to solve it is through discretization, which is equivalent to taking finite bins in $\alpha$. We will use this method here to find the KL eigenmodes that maximize the signal content for galaxy clustering.

    To do so, we have considered a specific strawman photometric survey targeting a sample of red galaxies, characterized by their higher bias and better photo-$z$ uncertainties than their blue counterparts (making them better suited for clustering analyses). The sample we consider is compatible with what could be observed by the Large Synoptic Survey Telescope \cite{2009arXiv0912.0201L}, characterized by the redshift distribution shown in the left panel of Fig. \ref{fig:nz_gc} (full details can be found in \cite{2015ApJ...814..145A}). We assume a photo-$z$ uncertainty of $\sigma_z=0.02\,(1+z)$ and split the sample into redshift bins in photo-$z$ space with $z_{\rm ph}>0.5$ and a width given by the photo-$z$ uncertainty at the bin centre. The auto-power spectra for our set of 23 bins are shown in the right panel of Fig. \ref{fig:nz_gc}. The large overlap between bins implies that a choice of thinner slices is unlikely to unveil significantly more information, and we have verified that the results shown below do not change after doubling the number of bins. All power spectra were computed using a modified version of the code presented in \cite{2013JCAP...11..044D}.
    \begin{figure}
      \centering
      \includegraphics[width=0.49\textwidth]{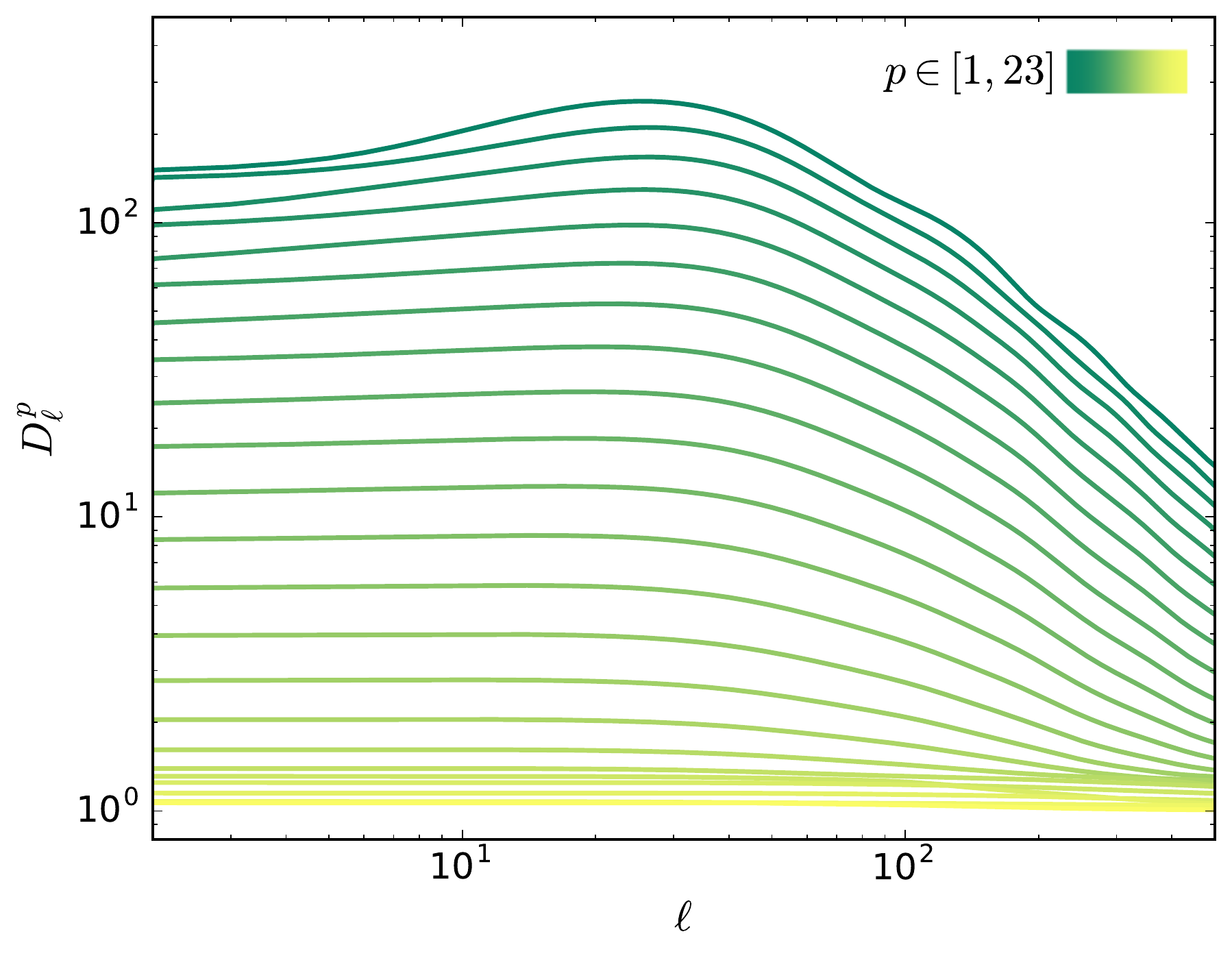}
      \caption{Power spectra of the KL eigenmodes for the strawman large-scale-structure survey. Unlike in the case of weak lensing (see Section \ref{ssec:results.wl}), a large number of eigenmodes are signal-dominated. This is due to the overall higher signal-to-noise ratio of galaxy clustering with respect to galaxy shear as well as to the smaller correlations between distant bins.}\label{fig:dp_gc}
    \end{figure}
    \begin{figure}
      \centering
      \includegraphics[width=0.49\textwidth]{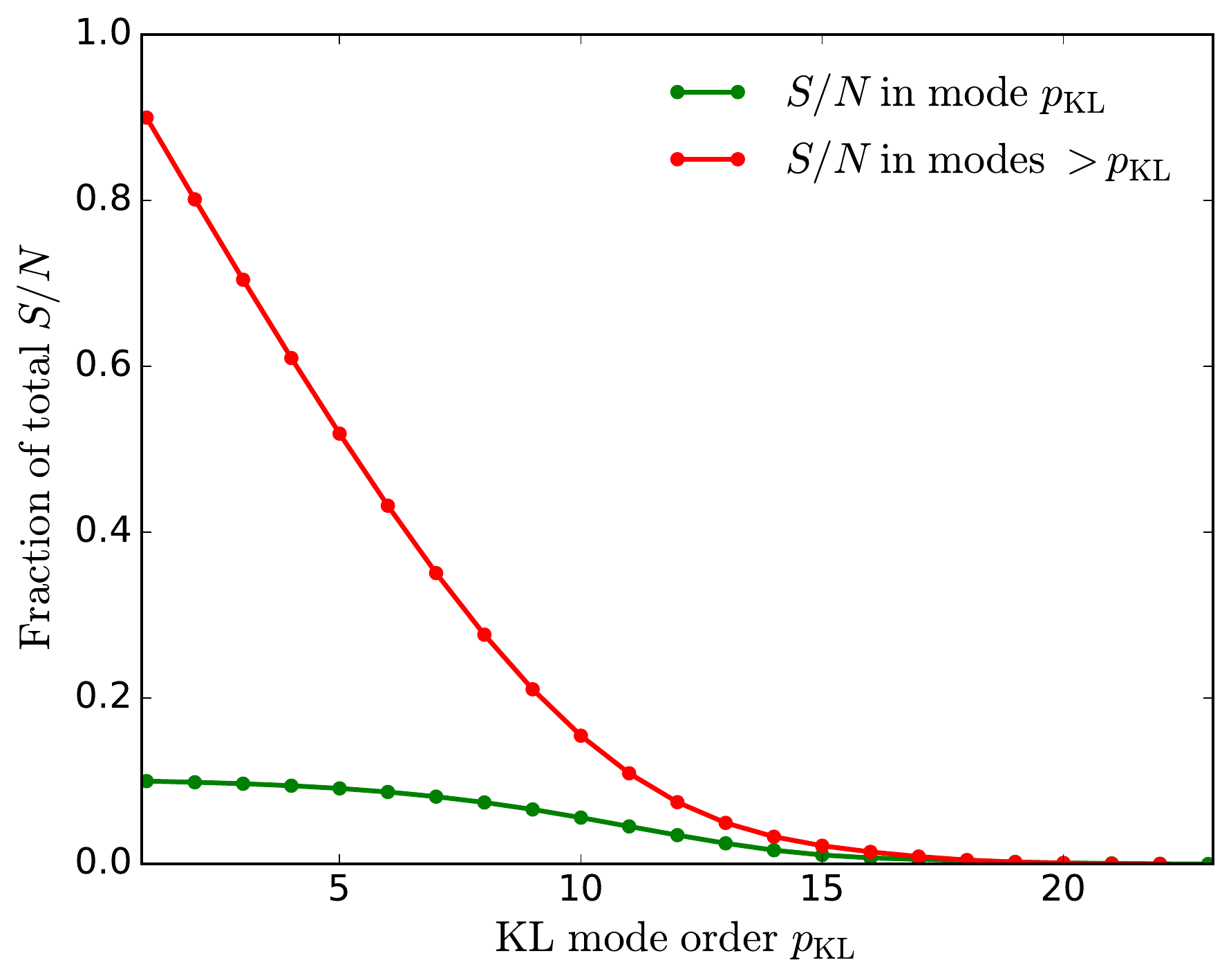}
      \includegraphics[width=0.49\textwidth]{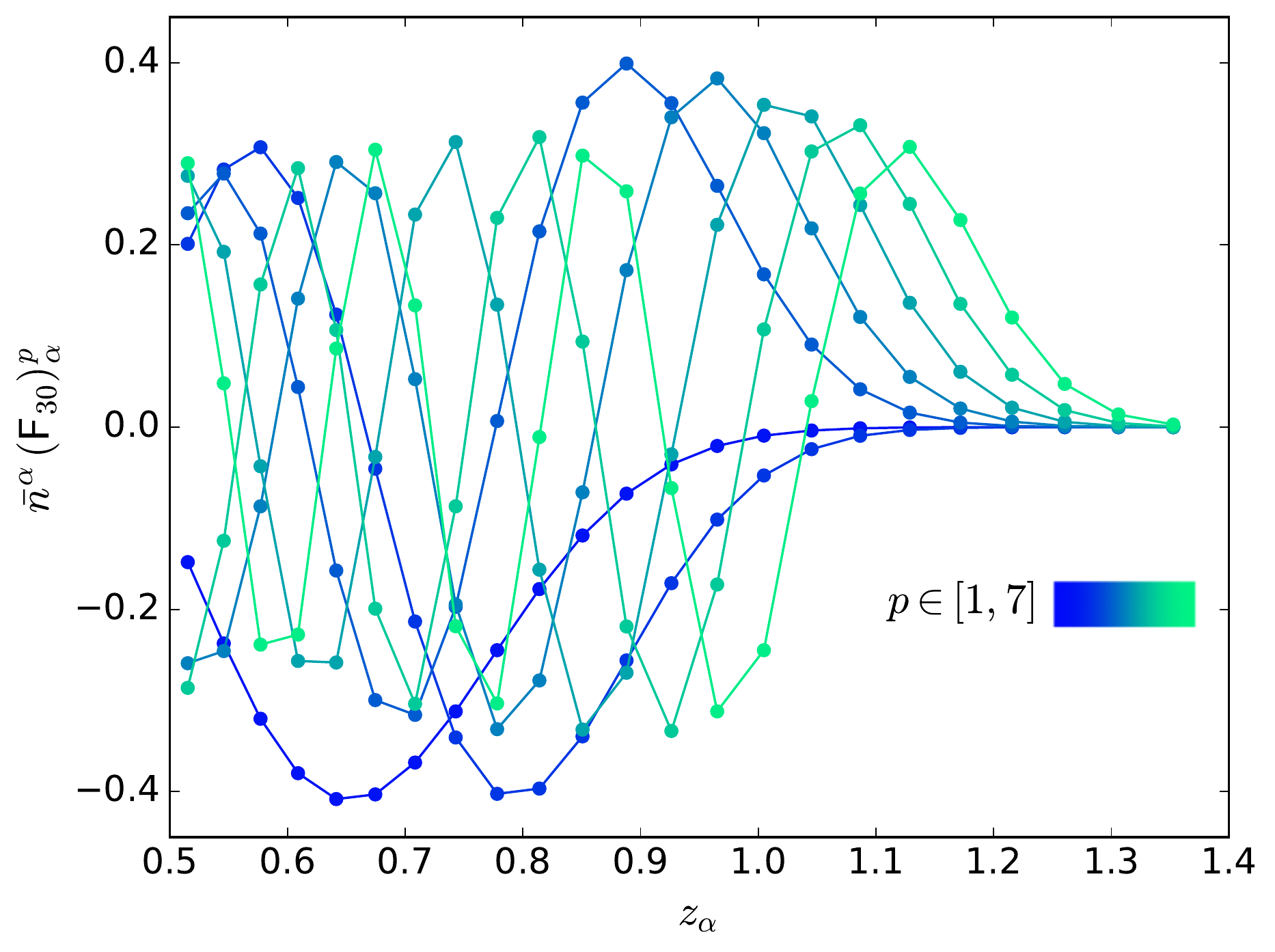}
      \caption{{\sl Top}: fraction of the total signal-to-noise ratio of the different KL eigenmodes for the strawman galaxy clustering survey. The bulk of the $S/N$ ($>90\%$) is encoded in the first 13 modes. {\sl Bottom}: the first 7 KL modes for $\ell=30$. The sinusoidal shape of the modes agrees with the expectation that, in the limit of $\sigma_z\rightarrow0$ and no background redshift dependence, the KL modes should be given by the spherical Bessel functions. }\label{fig:kl_gc}
    \end{figure}    

    Using the prescription described in Section \ref{sssec:method.klbasis.sn}, we find the KL eigenmodes and associated power spectra, and rank them according to their contribution to the total signal-to-noise ratio (defined here as the Fisher matrix element of the signal amplitude). The power spectra of the resulting KL modes are shown in Figure \ref{fig:dp_gc}. Unlike the case of weak lensing, explored in Section \ref{ssec:results.wl}, the information encoded in the galaxy overdensity is local in redshift, and thus the correlation between different bins decays rapidly with redshift separation. The signal-to-noise is therefore spread over $\sim15$ signal-dominated modes, and the noise-dominated modes can be thought of as the radial scales filtered out by the finite photo-$z$ uncertainty (as we mentioned in Section \ref{sssec:method.klbasis.sn}, the noise power spectrum gets mapped into $1$ under the KL transform). The relative contribution of each mode to the total signal-to-noise is shown in the top panel of Fig. \ref{fig:kl_gc}. 90\% of the total constraining power can be achieved by considering the first 13 eigenvectors. The form of the first 7 of these eigenvectors for $\ell=30$ are shown in the right panel of Fig. \ref{fig:kl_gc}. The eigenmodes are sinusoids with increasing frequencies, in agreement with the expectation that, in the limit of $\sigma_z\rightarrow0$ and no background redshift dependence, the KL decomposition is achieved by the spherical Bessel functions. A Fourier-Bessel decomposition is therefore probably a near-optimal analysis method for galaxy clustering, although the KL decomposition allows a more precise determination of the truly orthogonal radial modes.
    
  \subsection{Galaxy clustering - optimal basis for $f_{\rm NL}$}\label{ssec:results.fnl}
    It is expected that future large-scale photometric surveys will make the search for primordial non-Gaussianity one of their main science cases. This can be achieved by measuring the excess power on large scales caused by a non-zero value of $f_{\rm NL}$\footnote{The reader is referred to \cite{2004PhR...402..103B} for a thorough review of non-Gaussianity and a definition of $f_{\rm NL}$.} generates in the two-point statistics of biased tracers of the matter distribution \cite{2000ApJ...541...10M,2008PhRvD..77l3514D}. Since the signal is most relevant on large scales, we can expect the bulk of it to be concentrated in a small number of radial modes, which makes the general KL decomposition outlined in Section \ref{ssec:method.klbasis} an ideal analysis method. Similar approaches have been explored in the literature to devise optimal weights for spectroscopic galaxy surveys \cite{2017arXiv170205088M}.
    
\begin{figure}
      \centering
      \includegraphics[width=0.49\textwidth]{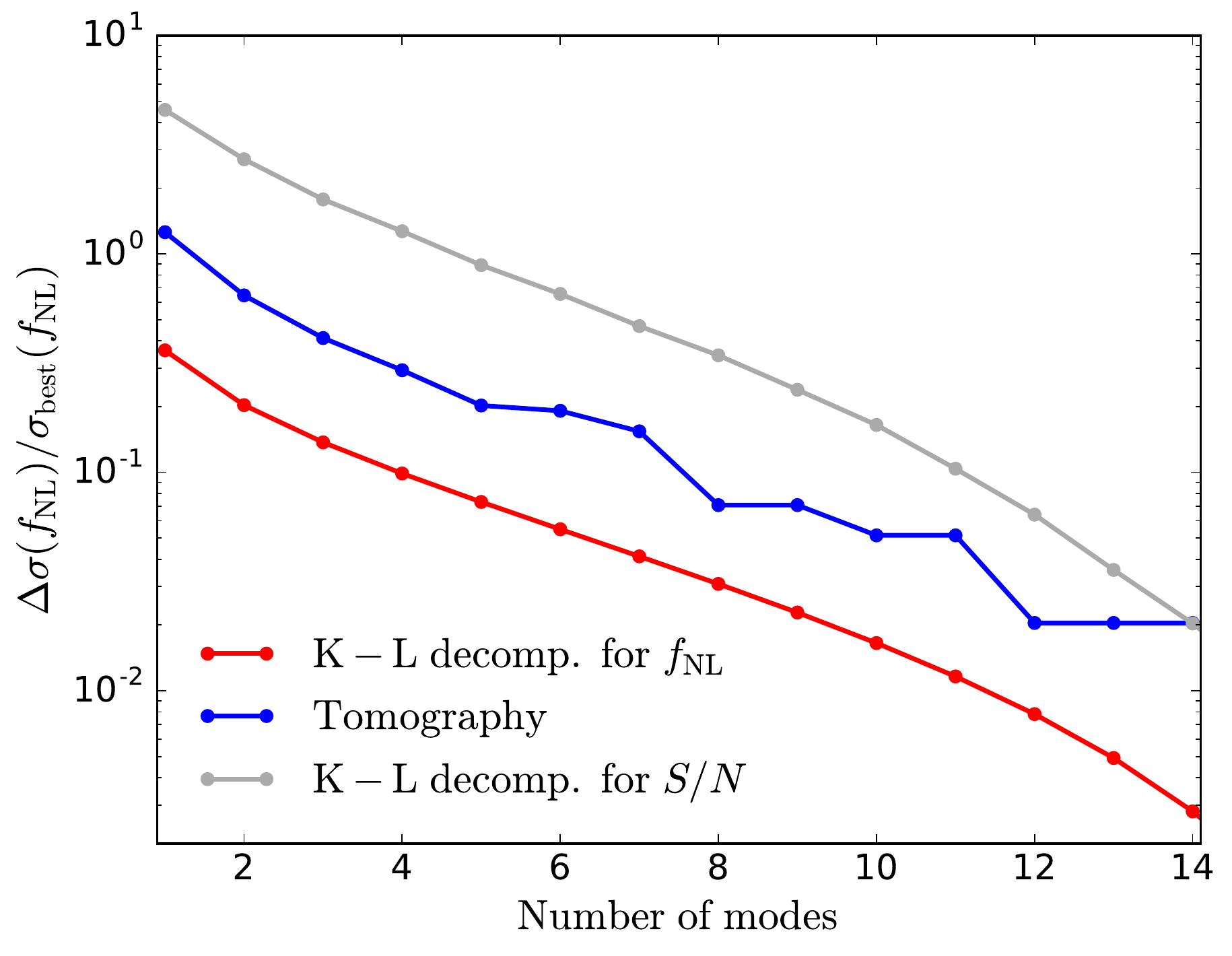}
      \caption{Excess uncertainty on $f_{\rm NL}$ with respect to the best achievable error on this parameter as a function of the number of modes included in the analysis for three different radial decomposition schemes: optimal KL modes for $f_{\rm NL}$ (red), optimal KL modes for the overall $S/N$ of the clustering signal (gray) and tomographic slicing into the corresponding number of bins of equal width (blue).}\label{fig:kl_fnl}
    \end{figure}

    We again consider the red galaxy sample used in the previous section, but now estimate the KL basis of eigenmodes that optimize the information content on $f_{\rm NL}$ instead of the overall signal amplitude. I.e. we solve the generalized eigenvalue problem in Eq. \ref{eq:kl_general} where $\theta=f_{\rm NL}$. We compare the performance of this basis with other choices of radial modes as follows: for a given number of modes, we estimate the associated uncertainty on $f_{\rm NL}$, $\sigma(f_{\rm NL})$, by summing the contributions to the corresponding Fisher matrix element of those modes, and compute the excess of $\sigma(f_{\rm NL})$ with respect to the best achievable constraint $\sigma_{\rm best}(f_{\rm NL})$. The results are shown in Fig. \ref{fig:kl_fnl} for three choices of radial functions:
    \begin{itemize}
      \item The KL eigenbasis resulting from optimizing the information content on $f_{\rm NL}$ discussed in this section (in red).
      \item The KL eigenbasis resulting from optimizing the overall signal-to-noise of the galaxy clustering signal, as discussed in the previous section (in gray).
      \item Photo-$z$ tomography: the result of dividing the galaxy sample into a number of top-hat photo-$z$ bins of equal width (in blue).
    \end{itemize}
    As demonstrated by this figure, for a fixed number of modes the optimal KL basis always outperforms any other data compression prescription. In particular, the constraints on $f_{\rm NL}$ are only degraded by $\sim30\%$ when considering only the first principal eigenmode, and almost 90\% of the total constraining power is contained in the first three. Interestingly, a naive tomographic approach achieves the same uncertainty on $f_{\rm NL}$ with a smaller number of modes (redshift bins) than the KL eigenbasis for the $S/N$. However, since the tomographic bins are not orthogonal, unlike the KL modes, for a fixed $\sigma(f_{\rm NL})$ both KL bases typically outperform the tomographic approach in terms of the size of the associated power spectrum. In any case, this example serves to stress the fact that the optimal radial basis in terms of overal $S/N$ is not necessary optimal in terms of final constraints for cosmological parameters that depend on specific features of the power spectrum.
    
  \subsection{Galaxy clustering - magnification bias}\label{ssec:results.mag}  
    \begin{figure}
      \centering
      \includegraphics[width=0.49\textwidth]{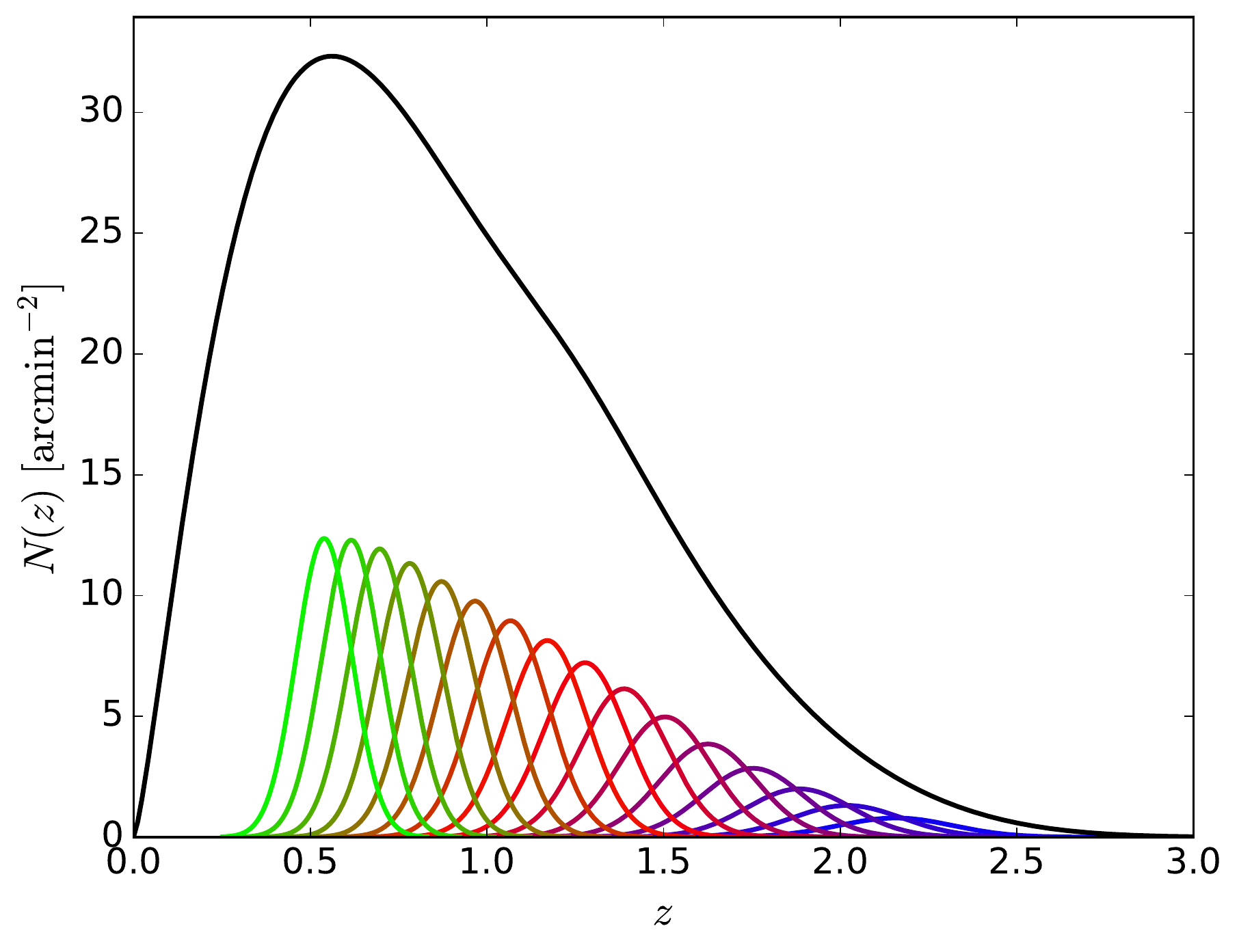}
      \caption{Redshift distribution and bins considered for the KL analysis of a strawman lensing survey (Section \ref{ssec:results.wl}) and for the extraction of the magnification bias signal from galaxy clustering (Section \ref{ssec:results.mag}).}\label{fig:nz_mb}
    \end{figure}
    Gravitational lensing of the observed galaxy positions alters their clustering pattern. This appears as an extra term in the galaxy clustering transfer function (Eq. \ref{eq:tgc_dr}):
    \begin{align}\nonumber
      &\Delta^{M,\alpha}_\ell(k)=-2\ell(\ell+1)\int\,d\chi W^{M,\alpha}(\chi)\frac{j_\ell(k\chi)}{k^2a(\chi)}\sqrt{P(k,z(\chi))},\\
      &W^{M,\alpha}(\chi)=\frac{3H_0^2\Omega_M}{2}\int_{z(\chi)}^\infty dz'\,\phi^\alpha(z')\frac{2-5\,s}{2}\,\frac{\chi(z')-\chi}{\chi(z')\chi},
    \end{align}
    where $s$ is the tilt in the number counts of sources as a function of magnitude limit. This effect, commonly labeled ``magnification bias'' \cite{1967ApJ...147...61G,2000ApJ...537L..77M,2008PhRvD..77b3512L}, can be used as an alternative measurement of gravitational lensing, through galaxy positions instead of shapes. The contribution of the magnification term is, however, weak in comparison with the density and RSD terms (Eq. \ref{eq:tgc_dr}), and therefore its measurement can be hampered by the cosmic variance contribution of these terms.
    
    \begin{figure}
      \centering
      \includegraphics[width=0.49\textwidth]{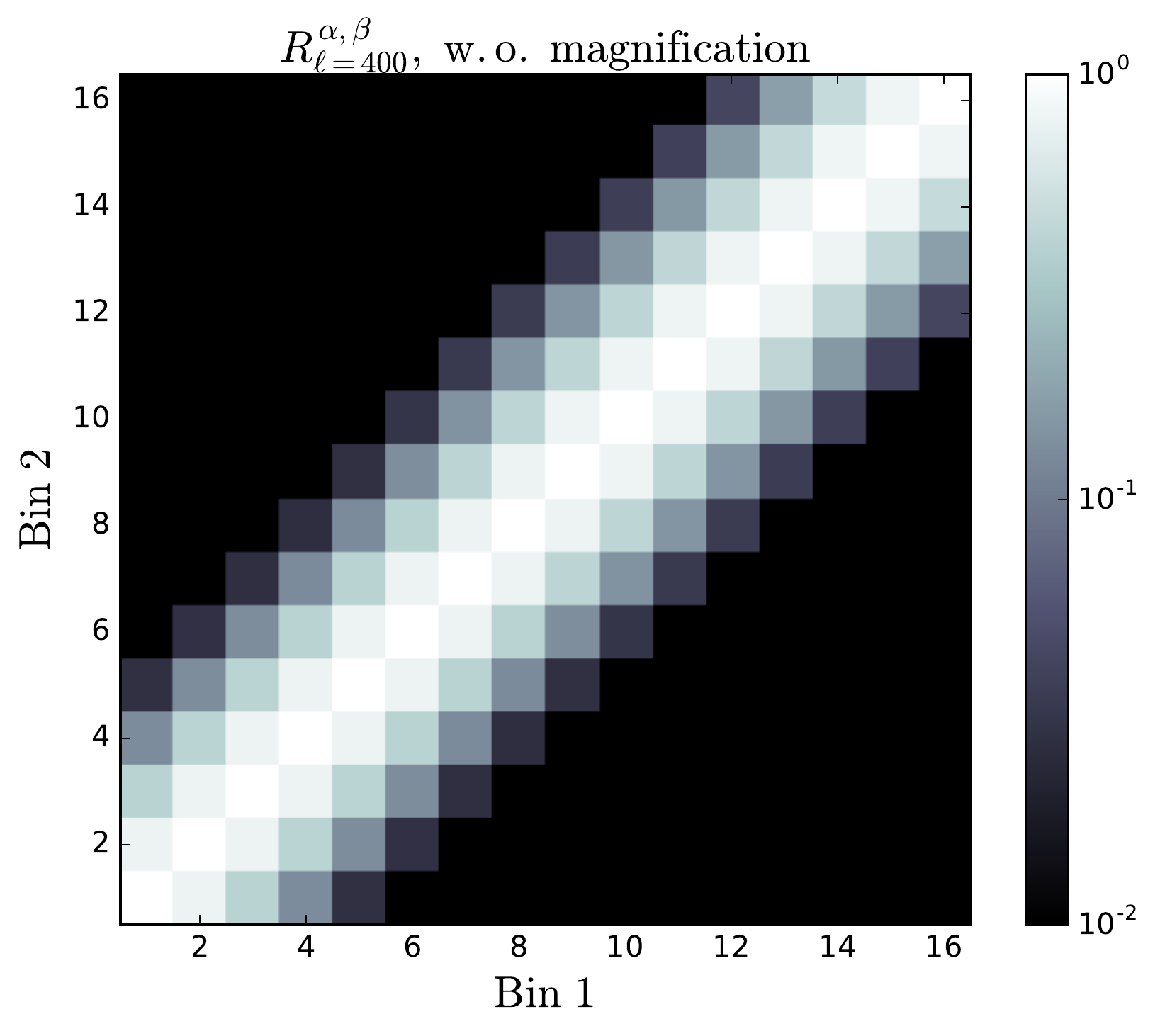}
      \includegraphics[width=0.49\textwidth]{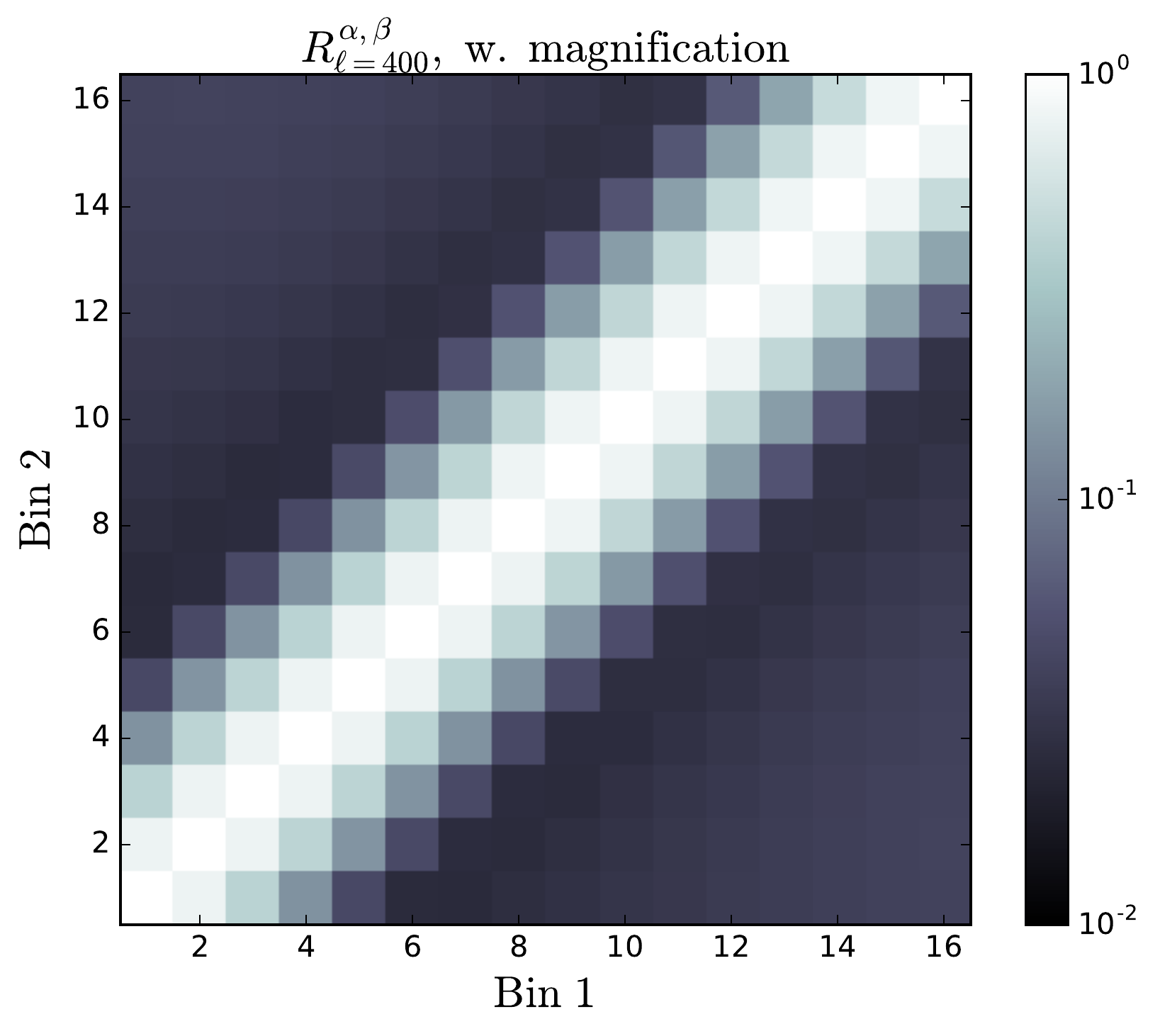}
      \caption{Correlation coefficient between the galaxy overdensities measured in the 16 redshift bins shown in Fig. \ref{fig:nz_mb}. The top panel shows the contributions of the true matter overdensity and redshift-space distortions alone. In this case the correlations between neighbouring bins are mostly caused by the overlap in redshift between them, and decays quickly with bin separation. The bottom panel then adds the contribution from lensing magnification, which generates a significant correlation between distant bins.}\label{fig:r_ij_wm}
    \end{figure}
    \begin{figure}
      \centering
      \includegraphics[width=0.49\textwidth]{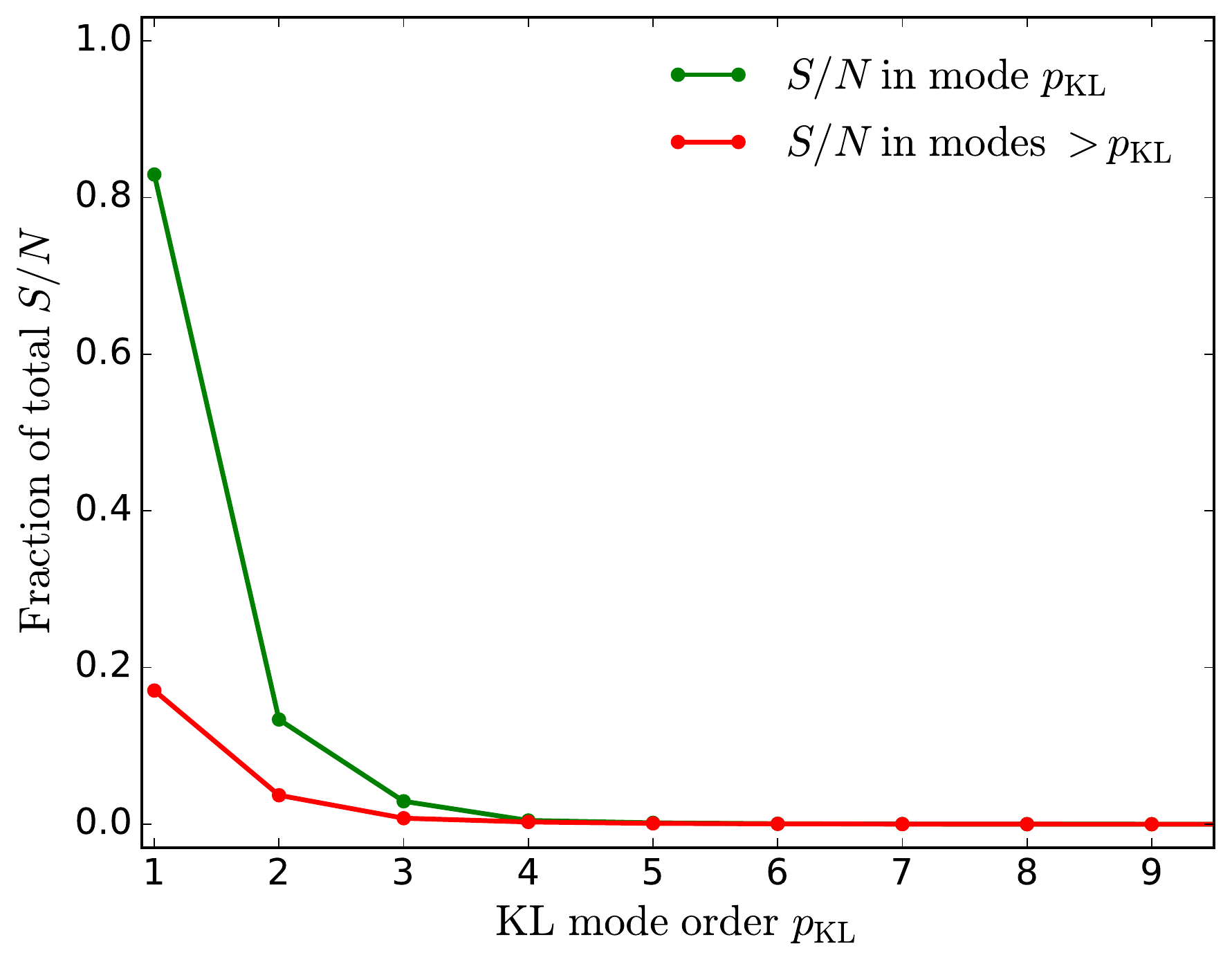}
      \caption{Fraction of the total signal-to-noise ratio of the magnification bias effect encoded in each KL eigenmode (green) as well as the cumulative information contained by all higher-order modes (red). The principal eigenmode contains $\sim80\%$ of the signal, and the first three modes are enough to capture it completely in practice.}\label{fig:kl_mb}
    \end{figure}
    One can therefore think of the density and RSD terms as correlated contaminants of the magnification signal, and use the KL formalism described in Section \ref{sssec:method.klbasis.cr} to devise an optimal basis of radial eigenmodes containing the bulk of its signal-to-noise.
    
    To test this approach we consider, as in the previous section, an LSST-like survey.  Since lensing magnification is an integrated effect, it is less hampered by poor photo-$z$ uncertainties, and it is most easily measured by cross-correlating high-redshift and low-redshift data \cite{2005ApJ...633..589S,2009A&A...507..683H}. For this reason, in this case we consider a sample of blue galaxies, with inferior photo-$z$ errors but wider redshift support. Full details can be found in \cite{2015ApJ...814..145A}. In summary, we consider a sample with $\sim40$ objects per arcmin$^2$ with the redshift distribution shown in Figure \ref{fig:nz_mb}. We also approximate the photo-$z$ distributions as Gaussians with a scatter $\sigma_z=0.05(1+z)$, and divide the sample into 16 top-hat bins in photo-$z$ space with $z_{\rm ph}<0.5$ and widths given by the value of $\sigma_z$ at the bin center (again, we verified that our conclusions did not change after decreasing the width by a factor 2).
    
    A key property of the magnification bias effect is the fact that, since gravitational lensing is caused by the integrated matter distribution between source and observer, the magnification signals in widely separated redshift bins can be tightly correlated. This is shown explicitly in Figure \ref{fig:r_ij_wm}. The figure shows the correlation coefficients between the 16 redshift bins, defined as $R^{\alpha\beta}_\ell=C^{\alpha\beta}_\ell/\sqrt{C^{\alpha\alpha}_\ell C^{\beta\beta}_\ell}$, at $\ell=400$, with (bottom panel) and without (top panel) the magnification bias effect. Although the contribution of lensing magnification to the correlation between neighbouring bins is subdominant, it produces noticeable correlations between distant ones.
    
    This property is particularly interesting in the context of the KL decomposition: a signal that is tightly correlated across samples will contribute significantly only to a small set of eigenmodes. To explore this possibility, we follow the prescription outlined in Section \ref{sssec:method.klbasis.cr} for correlated contaminants. The contribution of each eigenmode to the total signal-to-noise of the magnification bias (in a Fisher-matrix sense) is shown in Figure \ref{fig:kl_mb}. As expected, most of the signal ($>80\%$) is contained in the first eigenvalue, with the practical totality of it concetrated in the first three modes.
    
    We finish this section by noting that this approach is similar to the ``nulling'' method of \cite{2011MNRAS.415.1681H}, and that an analogous treatment could be carried out on the cosmic shear field to separate the lensing and intrinsic alignment contributions \cite{2008A&A...488..829J}.

  \subsection{Weak lensing}\label{ssec:results.wl}
    \begin{figure}
      \centering
      \includegraphics[width=0.49\textwidth]{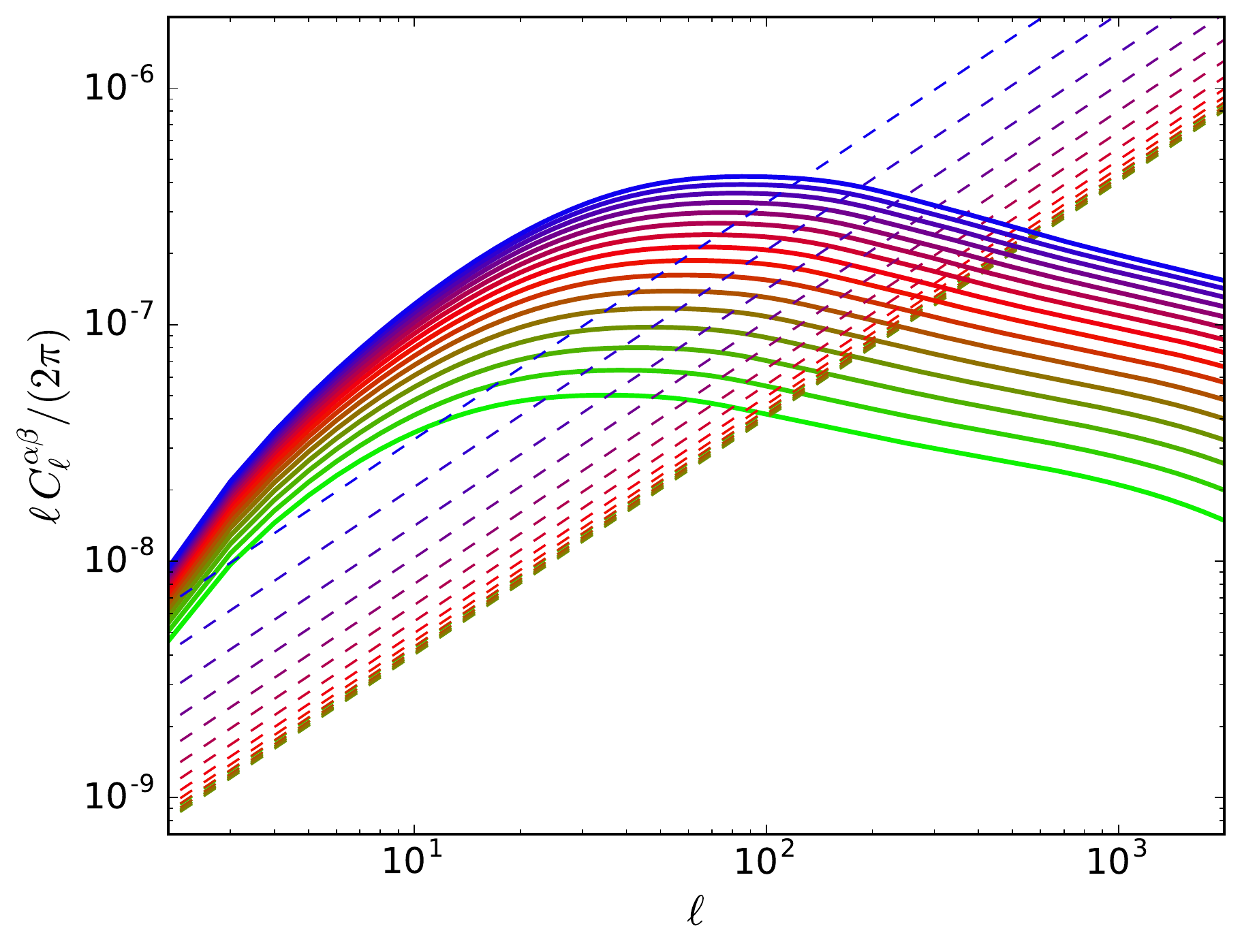}
      \caption{Shear auto-power spectra for the redshift bins shown in Fig. \ref{fig:nz_mb}. The signal and noise power spectra are shown as thick solid and thin dashed lines respectively.}\label{fig:nz_wl}
    \end{figure}
    \begin{figure}
      \centering
      \includegraphics[width=0.49\textwidth]{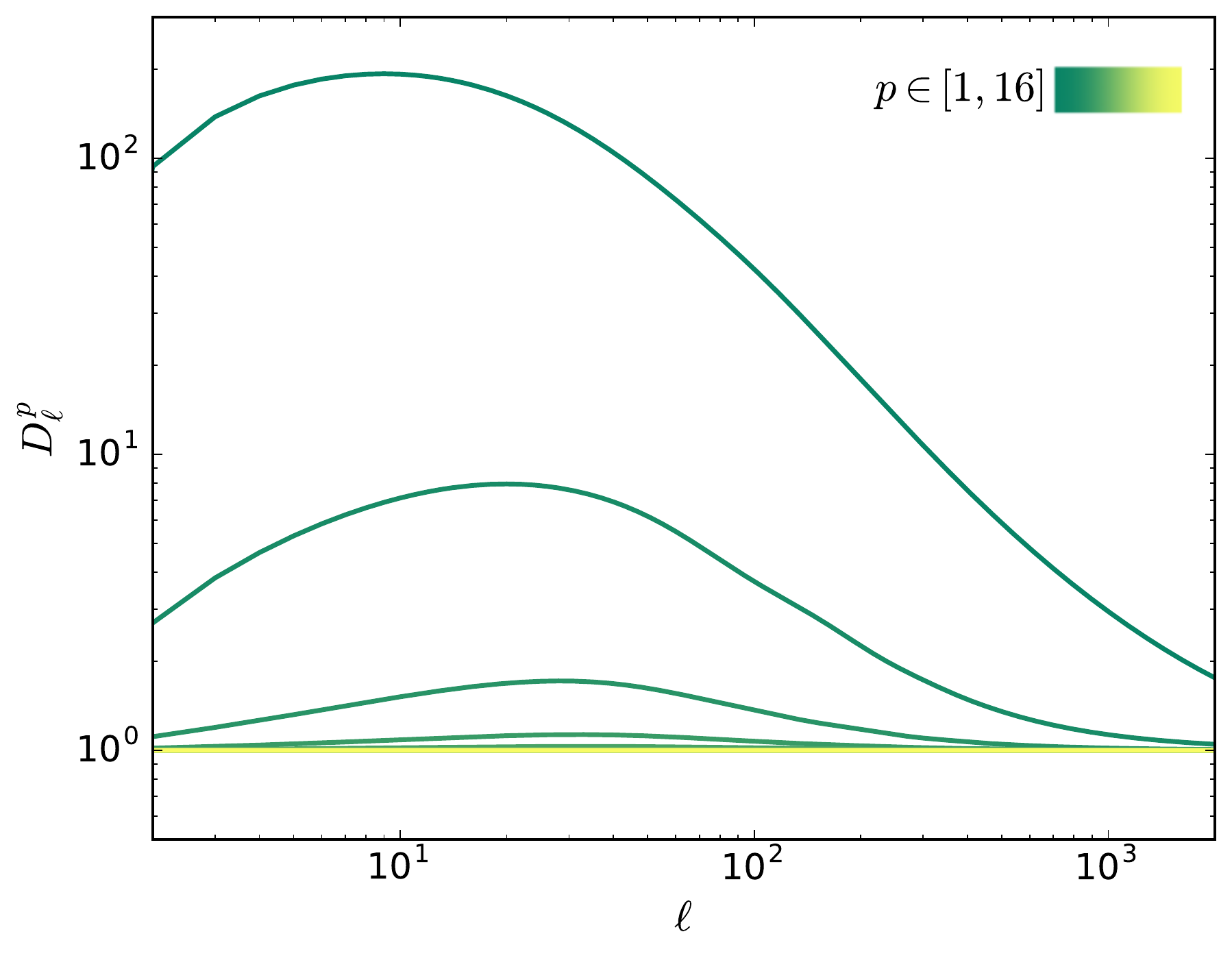}
      \caption{Power spectra of the KL eigenmodes for the strawman weak lensing survey. All but the first three modes are noise-dominated, and most of the information is encoded in the first mode.}\label{fig:dp_wl}
    \end{figure}

    The effects of gravitational lensing can be measured directly by studying the correlation it induces on the shapes and orientation of galaxy images. This effect, labeled ``cosmic shear'' is arguably the most promising observational probe for photometric redshift surveys, and therefore we will discuss the KL analysis of this signal in particular detail.
    
    As in the case of lensing magnification, and unlike the dominant galaxy clustering terms, the cosmic shear signal is correlated between widely separated redshift bins due to the integrated nature of gravitational lensing. Thus we can expect that a KL transform should be able to compress most of the signal to noise into a small set of radial eigenmodes. To quantify this we consider the same survey configuration used in Section \ref{ssec:results.mag}. The signal part of the cross-power spectrum between the cosmic shear measurements made in two different redshift shells is given again by Eq. \ref{eq:cl_generic}, where now the transfer functions $\Delta^{\alpha}_\ell$ take the form:
    \begin{align}\nonumber
      &\Delta^{\alpha}_\ell(k)\equiv\sqrt{\frac{(\ell+2)!}{(\ell-2)!}}\int d\chi\,W^\alpha(\chi)\frac{j_\ell(k\chi)}{k^2a(\chi)}\sqrt{P(k,z(\chi))},\\
      &W^\alpha(\chi)\equiv\frac{3H_0^2\Omega_M}{2}\int_{z(\chi)}^\infty dz\,\phi^\alpha(z')\frac{\chi(z')-\chi}{\chi(z')\chi}.
    \end{align}
    The noise power spectrum is white and simply given by the intrinsic ellipticity scatter weighed by the angular number density of sources in each redshift bin $\bar{n}^\alpha$:
    \begin{equation}
      N^{\alpha\beta}_\ell=\delta_{\alpha\beta}\frac{\sigma_\gamma^2}{\bar{n}^\alpha},
    \end{equation}
    with $\bar{n}^\alpha$ in units of ${\rm srad}^{-1}$ and $\sigma_\gamma=0.28$ \cite{2009arXiv0912.0201L}. The lensing auto-power spectra (both signal and noise) for these bins are shown in Figure \ref{fig:nz_wl}.
    
\begin{figure}
      \centering
      \includegraphics[width=0.49\textwidth]{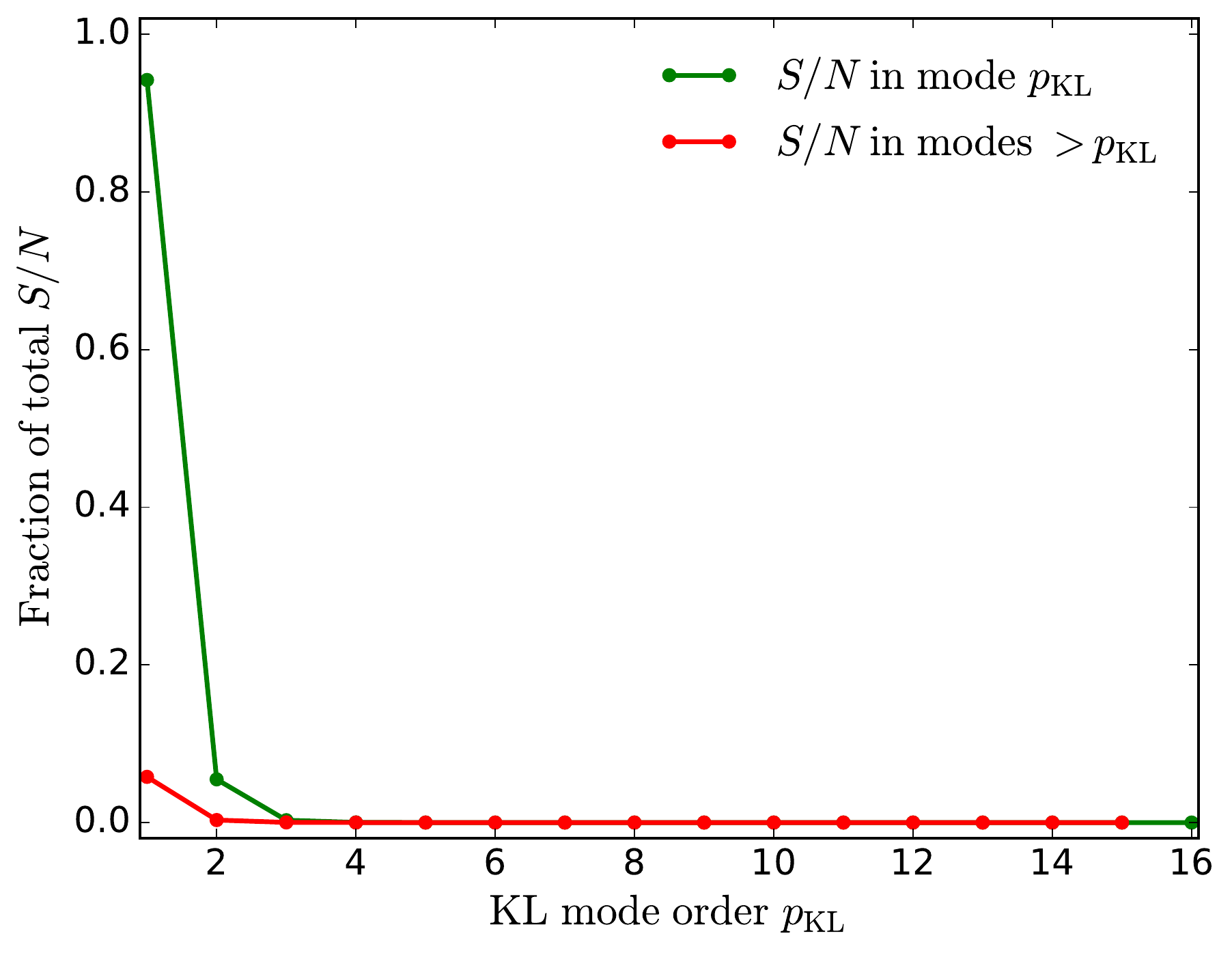}
      \includegraphics[width=0.49\textwidth]{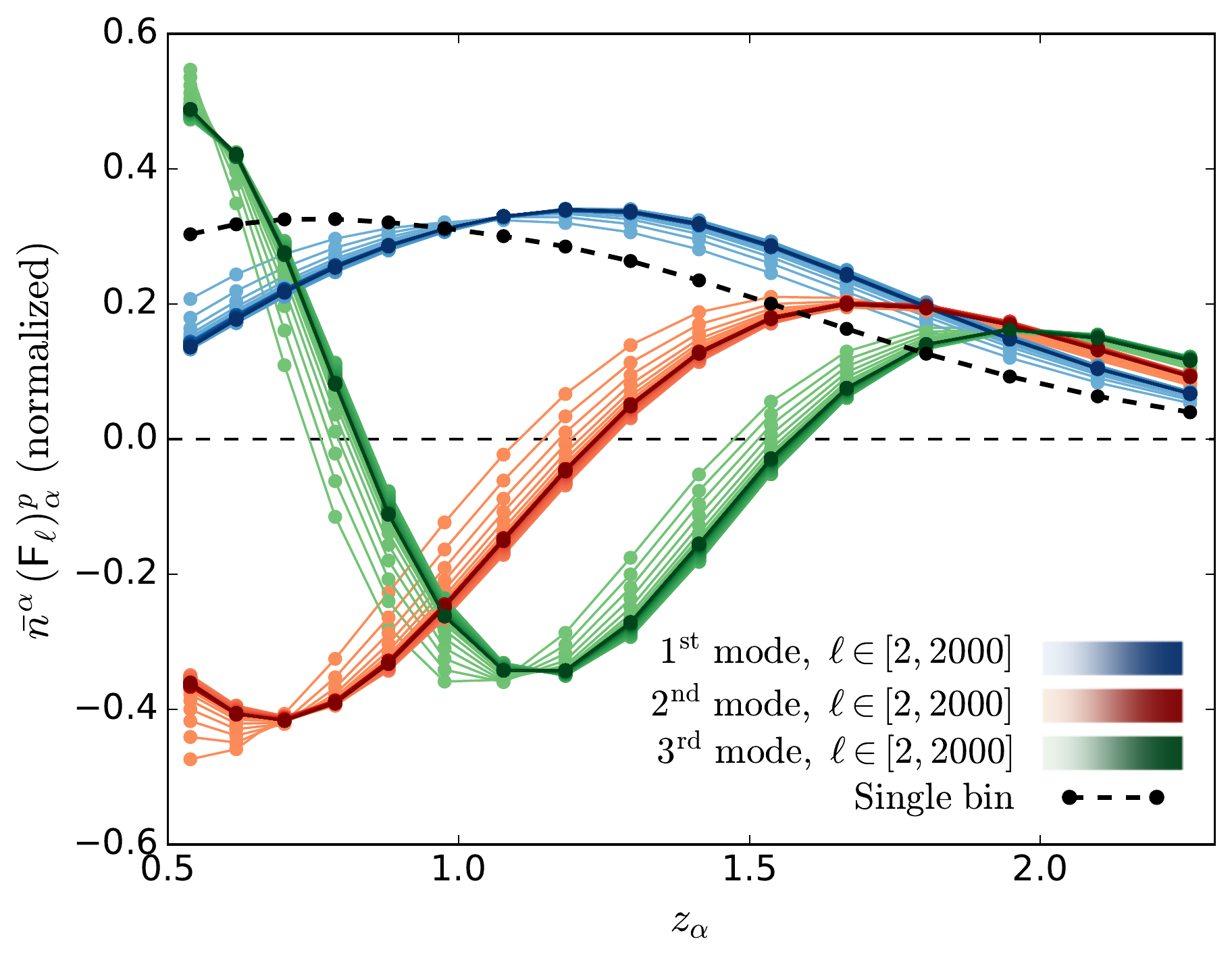}
      \caption{{\sl Top}: fraction of the total weak lensing signal-to-noise borne by each KL eigenmode (green) and the cumulative fraction contained in all higher-order modes (red). The first mode contains $\sim95\%$ of the signal, and the first three modes are enough to recover most of the information content. {\sl Bottom}: the first (blue), second (red) and third (green) KL eigenmodes of the strawman weak-lensing survey for different $\ell$. In all cases, the darkness of the line color increases with $\ell$. The redshift dependence of the modes stays roughly constant across $\ell$ and converges to a fixed shape for large $\ell$. The black dashed line corresponds to the weighting scheme associated with single-bin tomography (i.e. computing the signal integrated over the whole redshift range), which is similar to the weighting associated with the first principal eigenmode.}\label{fig:kl_wl}
    \end{figure}

    We compute the KL modes for this setup and rank them according to their contribution to the total lensing signal (in a Fisher matrix sense). The power spectra of the resulting set of modes are shown in Figure \ref{fig:dp_wl}. Comparing against Fig. \ref{fig:nz_wl} we can see that the KL decomposition effectively separates the signal-dominated and noise-dominated modes, with all modes $p>3$ dominated by noise. The fractional contribution of each mode to the total signal-to-noise is shown in the top panel of Figure \ref{fig:kl_wl}. Most of the signal ($\sim95\%$) is contained within a single mode, and the first two modes are able to recover more than $99\%$ of the total. The eigenvectors corresponding to the first three principal modes for different values of $\ell$ are shown in the right panel of the same figure. We observe that the eigenvectors preserve roughly the same shape for all $\ell$, and converge to the same shape at large $\ell$. The first eigenvector upweights the parts of the redshift range with the highest signal-to-noise, penalizing the low-$z$ regime due to its poor lensing signal and the high-$z$ bins due to their high shot noise. The second and third eigenmodes then recover part of this information by marginally upweighting these regions. The dashed black line in the same figure shows the weighting scheme associated with the measurement of the lensing signal integrated over a single bin encompassing the redshift range covered by the 16 bins in Fig. \ref{fig:nz_wl}. These weights are similar to the principal KL eigenmode, and thus the KL decomposition determines, broadly speaking, that the bulk of the signal-to-noise is mostly concentrated in the redshift-integrated signal, and extra information regarding the growth of structure can be picked up by up- or down-weighting the contributions in different sections of the redshift range.
    \begin{figure}
      \centering
      \includegraphics[width=0.49\textwidth]{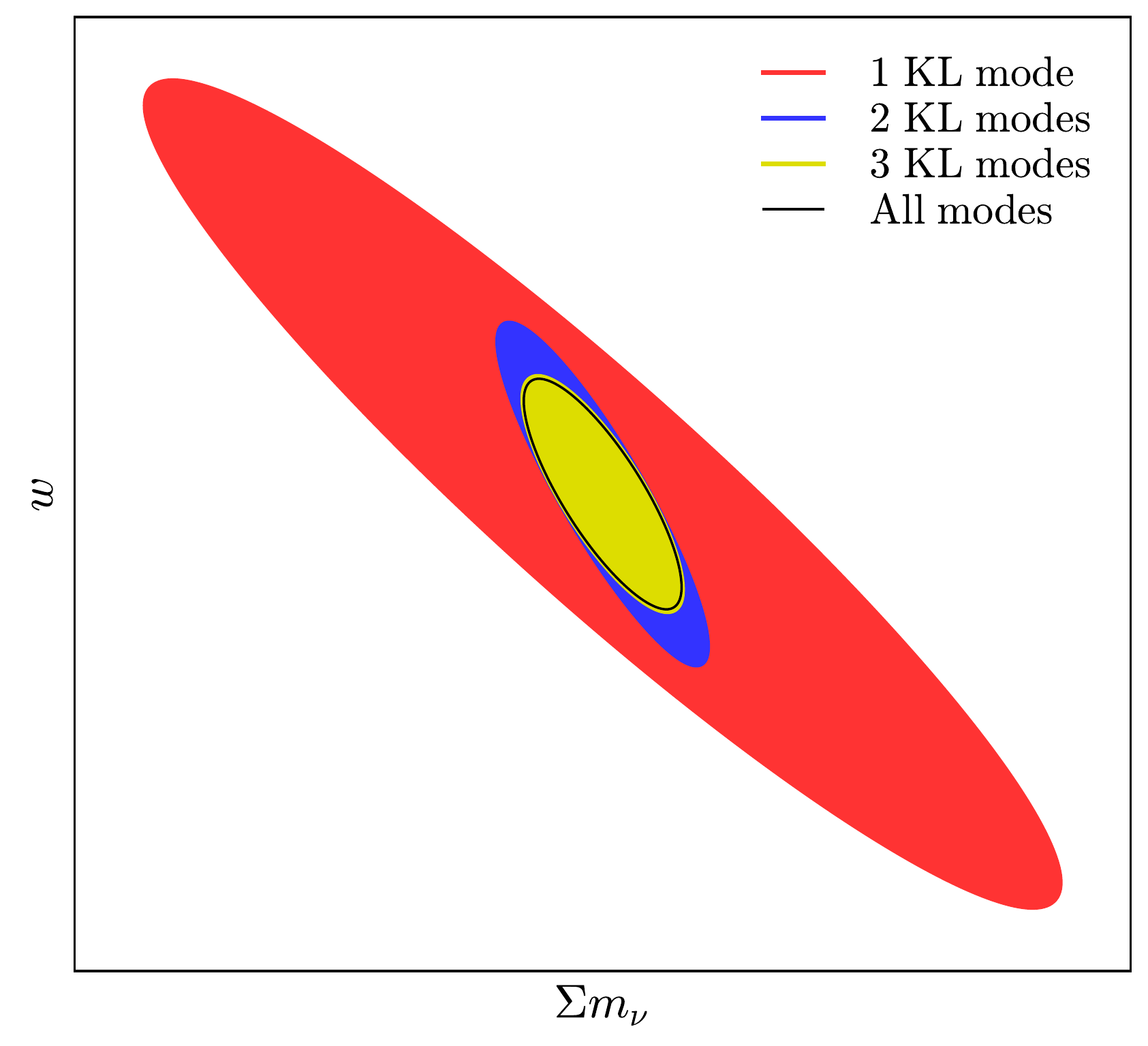}
      \caption{1$\sigma$ constraints on the dark energy equation of state $w$ and the sum of neutrino masses $\Sigma m_\nu$ achievable by analyzing the first (red), first two (blue) and first three (yellow) KL radial eigenmodes, compared with the best achievable constraints (solid black line). These constraints are marginalized over 7 other cosmological and nuisance parameters. Although the vast majority of the signal is encoded in the first mode, the next two modes are necessary in order to break the degeneracies between different parameters and recover optimal constraints.}\label{fig:fisher_wl}
    \end{figure}

    As we have discussed in the previous sections, the principal KL eigenmodes that optimize the recovery of the cosmological signal are not necessarily optimal in terms of encoding cosmological information, although it is plausible to expect so in general. In order study this further we have performed a Fisher-matrix forecast of the final constraints on cosmological parameters achievable by collecting the information encoded in the first $M$ principal eigenmodes, and compared them with the best possible constraints coming from the use of the full set of 16 redshift bins (or, equivalently, all of the KL eigenmodes). We do so following the approach described in Section 3 of \cite{2015ApJ...814..145A} and using, as observables, the corresponding set of KL modes $b^p_{\ell m}$. For these forecasts we considered a set of 9 parameters: the relative density of cold dark matter $\omega_c$, the relative contribution of baryons $\omega_b$, the normalized local expansion rate $h$, the amplitude $A_s$ and spectral index $n_s$ of primordial scalar perturbations, the sum of neutrino masses $\Sigma\,m_\nu$, the equation of state of dark energy $w$ and two parameters, $\log_{10}M_c$ and $\eta_b$, parametrizing the contribution of baryonic effects in the matter power spectrum as described in \cite{2015JCAP...12..049S}.
    
    Figure \ref{fig:fisher_wl} shows the results of this analysis in terms of $1\sigma$ contours in the $\Sigma m_\nu$-$w$ plane marginalized over all other parameters. The results are shown for the set of 1, 2 and 3 principal KL eigenmodes in red, blue and yellow respectively, while the best achievable constraints using all of the modes are shown as a solid black ellipse. We have removed the axis labels to focus the reader's attention on the relative improvement of the constraints with the number of modes. Even though the first eigenmode contains the vast majority of the lensing signal, as evidenced by the top panel of Fig. \ref{fig:kl_wl}, the extra information contained in the second and third eigenmodes is necessary in order to break the degeneracies between cosmological parameters. Three modes are however sufficient to recover the full constraining power with negligible loss of information.
    
    To finalize, we would like to emphasize the fact that, as shown in the bottom panel of Fig. \ref{fig:kl_wl}, the three principal eigenmodes preserve roughly the same shape as a function of multipole order, converging to the same curve for large $\ell$. An $\ell$-independent basis of radial functions would be advantageous from the point of view of data analysis since, for instance, the radial window functions associated with each mode (see Eq. \ref{eq:window_kl}) would only have to be calibrated once (independently of $\ell$). It is therefore interesting to explore the cosmological constraints achievable by using the radial functions associated with the KL eigenmodes at high $\ell$ \emph{for all $\ell$}, even though, for a fixed multipole order, the corresponding set of modes will not be exactly orthogonal. We have verified that, doing so for the first three KL eigenmodes, the final constraints on either $w$ or $\Sigma m_\nu$ degrade by less than $0.5\%$. This is a reasonable result, given the larger statistical weight of the small-scale (large-$\ell$) fluctuations.
    
\section{Discussion}\label{sec:discussion}
  Next-generation cosmological observations will gather their constraining power from a variety of observables, and will therefore have to deal with enormous data vectors. This will present a computational challenge, both from the point of view of likelihood evaluation and in the estimation of the covariance matrix. An efficient data compression scheme would be able to not only alleviate these problems, but also to separate the most significant and less contaminated modes in the data.
  
  In this article we have studied the problem of 3D data compression in the context of photometric redshift surveys, and presented a method, based on the KL transform, to derive a basis of orthogonal radial functions that optimally separate the data into uncorrelated modes, where optimality can be defined in terms of overall signal to noise ratio, information content on a particular cosmological parameter or separability between clean signal and contaminants. This basis is a general and natural extension of the well-know harmonic-Bessel (or Fourier-Bessel) decomposition of spherically-symmetric and translationally-invariant systems, adapted to the particular properties of the dataset under study. Even though the definition of this basis requires prior knowledge of some of these properties, including uncertain ones such as a model for the photo-$z$ distributions, once the radial eigenfunctions are selected, the analysis of the resulting data eigenmodes can proceed as usual, including any calibration of these properties. Thus, a suficiently well-educated model of the survey parameters should preserve the near-optimality of the associated eigenbasis, while not hampering the robustness of the analysis.

  We have shown that, for the study of galaxy clustering in an idealized spectroscopic survey, the optimal set of eigenmodes corresponds to the standard harmonic-Bessel basis, and that this would not be the case in the presence of redshift uncertainties, RSDs or in the analysis of weak lensing observables. For the study of galaxy clustering in a photometric redshift survey, we have shown that the KL basis that maximizes the recovery of the cosmological signal is Bessel-like, although more optimal compression schemes can be derived to optimize the measurement of individual cosmological parameters. In particular, in the case of $f_{\rm NL}$ we have shown that the bulk of the constraining power is concentrated in $\sim3$ radial modes. We have also extended the method to maximize the recovery of a particular signal in the presence of correlated contaminants, and shown that it could be used to simplify the measurement of the effect of lensing magnification as a subdominant contribution to the statistics of the galaxy distribution.
  
  In the case of cosmic shear measurements we have shown that, due to the integrated nature of the gravitational lensing effect, the bulk of the signal ($\sim95\%$) is concentrated in a single radial mode, qualitatively equivalent to the measurement of the weak lensing effect over the full redshift range of the survey. The next subdominant modes are however needed in order to break degeneracies between different parameters, and we have shown that three modes are enough to recover the best achievable cosmological constraints.
  
  Further work is needed in order to alleviate some of the practical shortcomings of the method: the KL decomposition is arguably less connected to real-space, directly observable quantities (although not less so than standard Fourier-space methods). Some of the usual methods for systematics calibration thus need to be adapted for a KL-based analysis, and this is particularly relevant for photo-$z$ systematics. In the case of weak lensing, however, we have shown that the shape of the radial eigenfunctions converges to the same curve on large multipole orders, and that the use of $\ell$-independent eigenfunctions would have a negligible impact on the final cosmological constraints. In this case, photo-$z$ calibration methods could be used in exactly the same manner as in the standard tomographic analysis.
  
  It is also worth emphasizing that, as is the case for the standard harmonic decomposition of fields defined on the sphere, the KL radial eigenmodes are no longer uncorrelated in the presence of an incomplete sky coverage, and a standard pseudo-$C_\ell$ analysis reveals non-zero coupling between different multipole orders $(\ell,\ell')$as well as different KL indices $(p,p')$ (see Appendix \ref{app:pcl} and \cite{2014MNRAS.442.1326K}). The impact of these correlations on the performance of the KL decomposition should be studied in more detail, and well-understood contaminant-deprojection techniques, implemented in standard power spectrum methods \cite{2017MNRAS.465.1847E}, should be adapted for this analysis.
  
  Finally, although we have explored the applicability of this method to independent galaxy clustering and weak lensing measurements, current and upcoming photometric redshift surveys will draw cosmological constraints from a joint analysis of both observables \cite{2017arXiv170609359K}. The direct application of this method to the joint data vector would in general produce eigenmodes that mix both signals. Alternatively a joint analysis of the KL modes of each observable, taken individually, could be performed, and the merits and drawbacks of each approach should be studied in detail.  
  
\section*{Acknowledgements}
  The author would like to thank Justin Alsing, Pedro Ferreira, Alan Heavens, Boris Leistedt, Jason McEwen, An\v{z}e Slosar and David Spergel for useful comments and discussions, and the Center for Computational Astrophysics, part of the Simons Foundation, for their hospitality. He also aknowledges support from the Science and Technology Facilities Council and the Leverhume and Beecroft Trusts. 
  
\bibliography{paper}

\appendix
\onecolumngrid
\section{Pseudo-$C_\ell$ estimation of the KL modes}\label{app:pcl}
  One of the standard methods to estimate the angular power spectrum of any two quantities in the cut sky is the so-called pseudo-$C_\ell$ estimator \cite{2002ApJ...567....2H}. This method can be directly applied to the two-point statistics of the KL eigenmodes, and reveals the correlation between radial modes generated by an incomplete sky coverage \cite{2014MNRAS.442.1326K}.
  
  The standard pseudo-$C_\ell$ method is based on computing the spherical harmonic coefficients of the masked field:
  \begin{equation}
    \hat{a}^\alpha_{\ell m}=\int d\nv\,a^\alpha(\nv)\,w^\alpha(\nv),
  \end{equation}
  where $w^\alpha$ is the weights map characterizing the mask of the field $a^\alpha$. One then estimates the power spectrum of this object by averaging over $m$ for each $\ell$:
  \begin{equation}
    \hat{C}^{\alpha\beta}_\ell\equiv\frac{\sum_m\hat{a}^\alpha_{\ell m}\hat{a}^{\beta *}_{\ell m}}{2\ell+1}.
  \end{equation}
  This object is then related to the true underlying power spectrum through a mode-coupling matrix $M^{\alpha\beta}_{\ell\ell'}$ such that
  \begin{equation}
    \left\langle\hat{C}^{\alpha\beta}_\ell\right\rangle=\sum_{\ell'}M^{\alpha\beta}_{\ell\ell'}C^{\alpha\beta}_{\ell'},\hspace{12pt}
    M^{\alpha\beta}_{\ell \ell'}\equiv\sum_{\ell''}\frac{(2\ell'+1)(2\ell''+1)}{4\pi}W^{\alpha\beta}_{\ell''}
    \left(
    \begin{array}{ccc}
      \ell & \ell' & \ell''\\
      0 & 0 & 0
    \end{array}
    \right)^2
  \end{equation}
  where the coupling matrix $M$ depends solely on the power spectrum of the masks $W^{\alpha\beta}_\ell\equiv(2\ell+1)^{-1}\sum_mw^\alpha_{\ell m}w^{\beta *}_{\ell m}$.
  
  The extension of this estimator to the power spectrum of the KL modes is straightforward: we project the masked harmonic coefficients $\hat{a}^\alpha$ over the KL eigenvectors ${\sf F}_\ell$ (i.e. $\hat{\bf b}_{\ell m}\equiv {\sf E}_\ell\circ\hat{\bf a}_{\ell m}$) and compute their power spectra by averaging over $m$. The resulting estimator takes the form $\hat{D}^p_\ell=\sum_{\ell'}M_{\ell\ell'}^{pp'}D^{p'}_{\ell'}$, where the new mode-coupling matrix is given by:
  \begin{equation}
    M^{pp'}_{\ell\ell'}\equiv \sum_{\alpha\beta}M^{\alpha\beta}_{\ell\ell'}\left[\sum_{\alpha'}({\sf F}_\ell)^p_\alpha({\sf N}^{-1})_{\alpha\alpha'}({\sf F}_{\ell'})^{p'}_{\alpha'}\right]\left[\sum_{\beta'}({\sf F}_\ell)^p_\beta({\sf N}^{-1})_{\beta\beta'}({\sf F}_{\ell'})^{p'}_{\beta'}\right]
    =M_{\ell\ell'}\left[\sum_{\alpha\beta}({\sf F}_\ell)^p_\alpha({\sf N}^{-1}_\ell)_{\alpha\beta}({\sf F}_{\ell'})^{p'}_\beta\right]^2
  \end{equation}
  where the second equality holds only if all the maps $a^\alpha_\ell$ share the same mask $w$. Note that, for full-sky coverage $M_{\ell\ell'}=\delta_{\ell\ell'}$, and using the orthonormality of ${\sf F}$ we recover $M^{pp'}_{\ell\ell'}=\delta_{\ell\ell'}\delta_{pp'}$.

\end{document}